\documentclass[a4paper,fleqn]{cas-dc}

\usepackage[numbers,compress]{natbib}
\usepackage{subcaption}
\usepackage{tikz}

\usetikzlibrary{shapes.geometric}
\usetikzlibrary{arrows}
\usetikzlibrary{arrows.meta}
\usetikzlibrary{positioning}
\usetikzlibrary{fit,tikzmark}
\usepackage{array}
\usepackage{colortbl}
\usepackage{placeins}
\usepackage[normalem]{ulem} 

\usepackage{expl3}
\expandafter\def\csname ver@l3regex.sty\endcsname{}

\newcommand{\myball}[1]{\textcircled{#1}}

\def\tsc#1{\csdef{#1}{\textsc{\lowercase{#1}}\xspace}}
\tsc{WGM}
\tsc{QE}

\definecolor{grey4}{HTML}{777777}
\definecolor{grey3}{HTML}{999999}
\definecolor{grey2}{HTML}{CCCCCC}
\definecolor{grey1}{HTML}{E6E6E6}
\definecolor{textcolor}{HTML}{FFFFFF}

\newcommand{\sreview}{\color{black}}
\newcommand{\ereview}{\color{black}}
\renewcommand{\sout}[1]{}

\tikzset{
  arr/.style={
    ultra thick
  },
}
 
\tikzset{
  arrlabel/.style={
    sloped,
    above
  },
}

\newcommand{\myboxcontent}[2]{\centering \textbf{#1} \\ #2}

\tikzset{
  mybox/.style={
    rectangle,
    rounded corners,
    draw=black, thick,
    inner sep=5pt,
  },
}

\newcommand{\Y}{\mathcal{Y}}
\newcommand{\R}{\mathbb{R}}

\begin{document}
\let\WriteBookmarks\relax
\def\floatpagepagefraction{1}
\def\textpagefraction{.001}

\shorttitle{Generalized Inverse Optimal Control and its Application in Biology}    

\shortauthors{Banga and Sager}  

\title [mode = title]{Generalized Inverse Optimal Control and its Application in Biology}  



%

\author[1]{Julio R. Banga}[orcid=0000-0002-4245-0320]

\cormark[1]


\ead{j.r.banga@csic.es}

\ead[url]{https://www.bangalab.org}


\affiliation[1]{organization={Computational Biology Lab, Misión Biológica de Galicia (MBG-CSIC),
Spanish National Research Council},
            city={Pontevedra},
          citysep={}, 
            postcode={36143}, 
            country={Spain}}

\affiliation[2]{organization={Department of Mathematics, Otto von Guericke University},
            city={Magdeburg},
          citysep={}, 
            postcode={39108}, 
            country={Germany}}

\affiliation[3]{organization={Max Planck Institute for Dynamics of Complex Technical Systems},
            city={Magdeburg},
          citysep={}, 
            postcode={39108}, 
            country={Germany}}

\author[2,3]{Sebastian Sager}[orcid=0000-0002-0283-9075]
\cormark[1]

\ead{sager@ovgu.de}

\ead[url]{https://mathopt.de}


\cortext[1]{Corresponding author}



\begin{abstract}
Living organisms exhibit remarkable adaptations across all scales, from molecules to ecosystems. 
We believe that many of these adaptations correspond to optimal solutions driven by evolution, training, and underlying physical and chemical laws and constraints.
While some argue against such optimality principles due to their potential ambiguity, we propose generalized inverse optimal control to infer them directly from data. This \sreview comprehensive \ereview approach incorporates multi-criteria optimality, nestedness of objective functions on different scales, the presence of active constraints, the possibility of switches of optimality principles during the observed time horizon, maximization of robustness and minimization of time as important special cases, as well as uncertainties involved with the mathematical modeling of biological systems.
This data-driven approach ensures that optimality principles are not merely theoretical constructs but are firmly rooted in experimental observations. The inferred principles can \sreview also \ereview be used in forward optimal control to predict and manipulate biological systems, with possible applications in bio-medicine, biotechnology, and agriculture. 
As discussed and illustrated, the well-posed problem formulation and the inference are challenging and require a substantial interdisciplinary effort in the development of theory and robust numerical methods. 
\end{abstract}



\begin{keywords}
 inverse optimal control \sep optimality principles \sep biology \sep optimization \sep mathematical modeling \sep inverse problems \sep bioengineering
\end{keywords}

\maketitle


\section{Introduction} \label{sec_introduction}

Optimality principles are a core concept in various scientific fields,  representing the idea that systems tend to evolve or function to achieve the best outcome under specific constraints. 
This concept serves two purposes: aiding in decision-making (e.g., in engineering or operations research) and facilitating the understanding of fundamental concepts and natural laws (e.g., in mathematics, physics, and chemistry).
When defining ``optimal'', we predominantly rely on a formal mathematical definition. In \textbf{mathematical optimization}, the focus lies on determining the values of variables $y$ such that an \textit{objective function} $\phi(y)$ is minimal among all \textit{feasible} choices $y \in \Y$ with a given \textit{feasible set} $\Y$. It is worth noting that maximization problems can be readily reformulated by seeking the minimum of $- \phi(y)$. In the context of time-dependent processes, \textbf{optimal control} involves influencing the dynamic system in an optimal way.
There are many specifications possible and necessary. 
In the interest of a non-technical presentation, we leave an introduction of mathematical concepts and models for particular real-world processes until Section~\ref{sec_gIOC}. For now, let us proceed with the intuitive understanding of optimality as the best among all available alternatives, which we aim to use for decision-making or to enhance understanding.   

Optimality principles have had decisive influence across various domains since ancient times. Among the earliest instances are those in \textbf{geometry}, manifested in problems aimed at finding extremal values (minimum or maximum) of quantities like distance, area, or volume. Probably the oldest known example is the isoperimetric problem, also known as Dido's problem \cite{Bandle2017-ur}, which seeks to ascertain the shape of a closed plane curve of fixed length that encloses the maximum possible area. It finds its origins in the mythical tale of the foundation of Carthage by the Phoenician Queen Dido, who was promised in mockery as much land as could be enclosed by a bull's hide. According to myth, she cut the hide into a long, thin strip and used it to bound the maximum possible area with a circle. Early mathematicians like Zenodorus in the second century B.C., relied on intuitive arguments to find the solution. 
A rigorous proof was given by Weierstrass in 1927, building upon a previous trial by Steiner from the 19th century \cite{Nahin2021-lt}.
Another old extremal problem in geometry is the geodesic problem, which involves finding the shortest path between two fixed points on a surface, with motion constrained to the surface. Even though ancient Greek mathematicians were aware of the solution of this problem in the plane, a rigorous proof did not emerge until the 18th century \cite{Nahin2021-lt}.

In physics and chemistry, optimality plays a fundamental role in characterizing natural processes.
For example, in \textbf{physics}, \textit{Fermat’s principle} states that the path followed by a ray between two points is the one that minimizes travel time. Over the centuries, this principle evolved into the \textit{Principle of Least Action}, notably expanded upon by Leibniz, Euler, Lagrange, and Hamilton. For a historical overview, refer to \cite{Ramm2011-me}. With Lagrange and Hamilton's contributions, we arrived at the general principle that -- for each particle in a conservative or non-conservative system -- the action taken from its initial position to its final position is optimized. The Lagrangian approach applies to all current physical theories, including general and special relativity to quantum mechanics and even string theory. A comprehensive introduction to this topic can be found in Chapter 19 of \cite{feynmanLecturesV2}. Moreover, in \textbf{chemistry}, the maximization of entropy or minimization of potential energy stand as key concepts. A notable instance is observed in protein folding. Numerous computational methods for predicting protein structure from its amino acid sequence rely on the principle that the native state of a protein possesses the lowest free energy, therefore representing the most stable configuration \cite{Anfinsen1973-qw}. Optimality principles in physics and chemistry are extensively documented, and for comprehensive surveys, we refer to \cite{Nesbet2002-uz,Lanczos1986-ih,Rektorys2012-xz}. 

In \textbf{economics} and \textbf{psychology} optimality holds crucial importance across various schools of thought. The \textit{homo oeconomicus} concept, presumed to pursue subjectively defined goals optimally, forms the foundation of many economic theories. Additionally, optimality serves as a methodological tool, e.g., in the analysis of tax impact via related optimal control problems \cite{leonard1992optimal} or in using optimal solutions as objective performance measures in experimental studies on complex problem solving \cite{Sager2011c}. For a critical discussion of optimality's role in economics, psychology, and other sciences see \cite{Schoemaker1991-nt} .
Moreover, driven by humanity's desire for improvement in all aspects of modern living, optimization principles are integral to decision support in all \textbf{engineering and operations research sciences} . Notably, in the realm of \textbf{artificial intelligence and machine learning} the persuasive technology of optimization has extended its reach into various domains of data-rich science and modern life.

Before exploring into the central theme of this article, which explores optimality principles in \textbf{biology}, it is important  to establish the conceptual distinction between \textbf{forward} and \textbf{inverse problems}.
In scientific inquiry, an inverse problem entails deducing the causal factors that led to a set of observations. In simpler terms, inferring a mathematical description (and hence of interpretable scientific explanation) based on observed data.
Forward and inverse simulations are well-recognized examples of this distinction.
\\
\vspace*{-0.4cm}
\begin{center}    
\begin{tikzpicture}
    \node[draw, rectangle, rounded corners, inner sep=5pt, fill=grey1] (model) at (0,0) {Model};
    \node[draw, rectangle, rounded corners, inner sep=5pt, fill=grey1] (observations) at (4,0) {Observations};
    \draw[->, thick, bend left] (model.north) to node[midway,above] {Forward: prediction} (observations.north);
    \draw[->, thick, bend left] (observations.south) to node[midway,below] {Inverse: inference} (model.south);
\end{tikzpicture}
\end{center}    
\vspace*{-0.4cm}

\noindent
In this context, ``model'' refers to a (mathematical) simulation model, which could be a set of differential equations. It is worth noting that the (regression) task of identifying functional representations and/or estimating model parameters for the mathematical model use optimization methods.
To illustrate, consider the instructive example of car driving: forward simulation could forecast material wear, while an inverse simulation might identify the cause of observed oscillations at the driver's seat.

We shall generalize this concept within the realm of optimization by interpreting ``model'' as an optimization model $\min \phi(y)$ subject to $y \in \Y$. With this interpretation, we can then make forward predictions for observations given $\phi$ and $\Y$. Conversely, the inverse optimization problem infers the unknown components of $\phi$ and $\Y$ from given observations.
The forward car driving scenario might involve determining a policy for accelerating, switching gears, and steering a car efficiently \sreview to \ereview navigate from $A$ to $B$ in minimal time, while adhering to constraints imposed by the track and engine speed. Conversely, the inverse optimization problem entails observing the car's position and velocity and determining whether the driver is prioritizing energy-minimal or time-minimal driving.

\begin{figure*}[ht!!!]
\centering
  \begin{subfigure}{0.48\linewidth}
  \includegraphics[width=\linewidth]{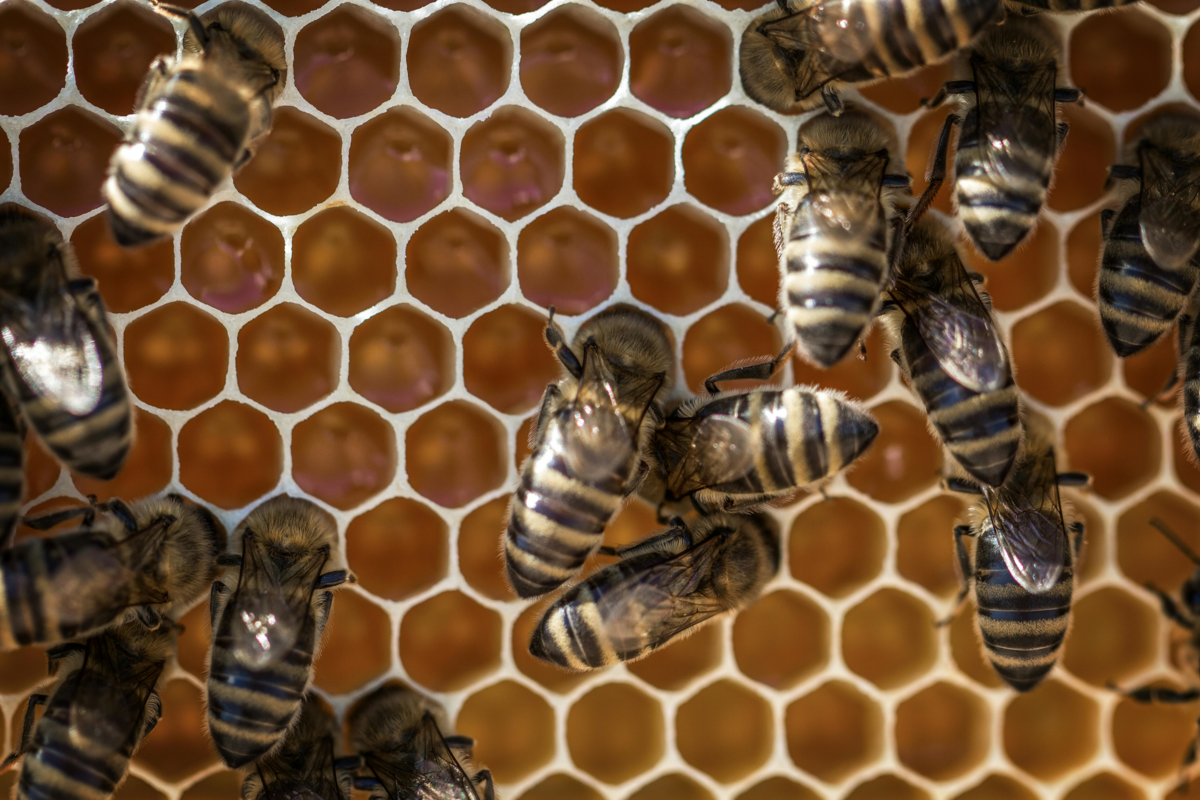}
  \caption{Hexagonal honey combs are optimal} \label{fig:subfigB}
  \end{subfigure}
  \begin{subfigure}{0.48\linewidth}
  \includegraphics[width=\linewidth]{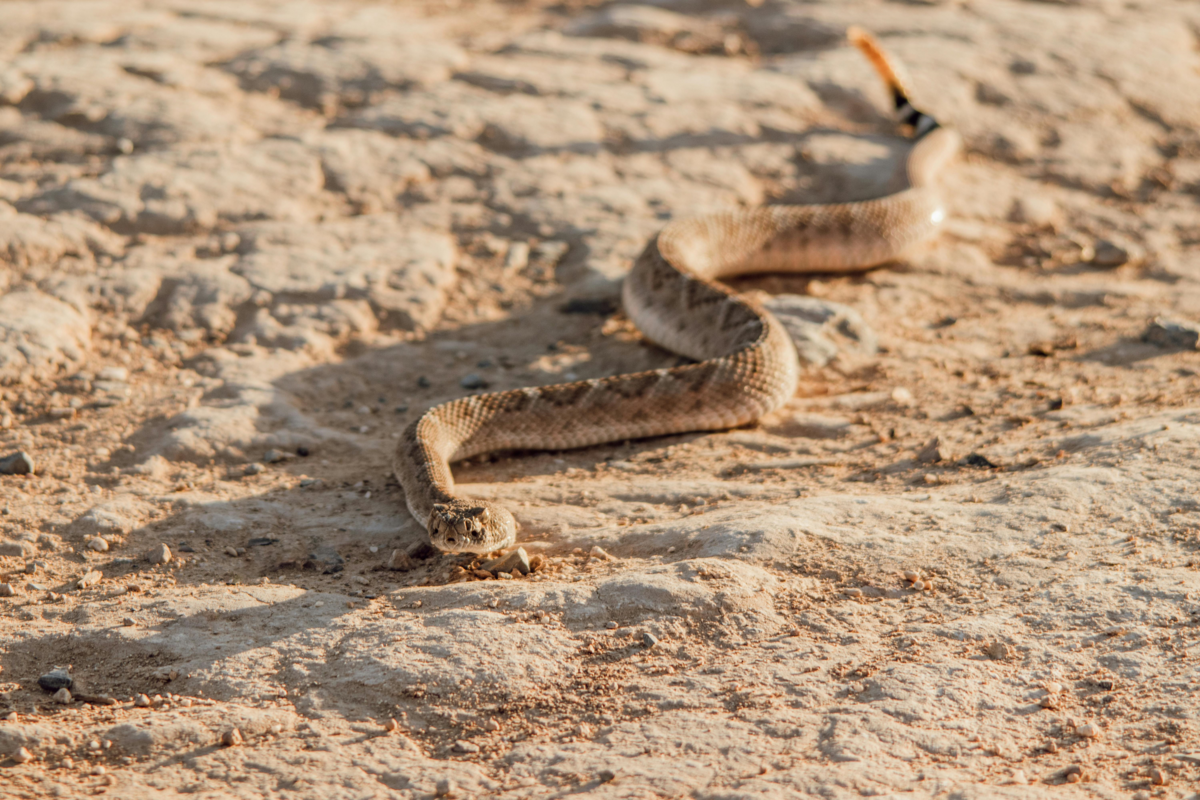}
\caption{Undulatory movements of snakes minimize drag forces} \label{fig:subfigC}
  \end{subfigure}
\caption{Examples of optimality principles observable in real life.
%
\eqref{fig:subfigB}: Already the Roman scholar Varro hypothesized that bees use a regular hexagonal lattice because it is the most efficient way to store honey while using the least wax. In 1999, Thomas Hales gave a mathematical proof for this.
\eqref{fig:subfigC}: Optimality principles have also been investigated for the movement of animals and humans. Shown is the example of undulatory movements of snakes, which were conjectured to minimize drag forces \cite{Astley2015}. 
%
}
\label{fig_optimalityExamples}
\end{figure*}




Our main goal is to provide an overview of the use of optimality principles in the biological sciences, focusing specifically on the cellular level. We discuss how these principles have been and could be used to explain various biological processes and systems. We highlight the main challenge unique to this field: unlike in previously mentioned contexts, the functions being optimized in biology are often unknown a priori. Simultaneously, it is highly desirable to have at least approximate knowledge of these functions whether for exploitation in an engineering context or for scientific insight. To address this issue and tackle several criticism often raised against the general consideration of optimality in biology within the literature, we present the \sreview \sout{novel} \ereview \textit{generalized inverse optimal control} framework. This framework builds upon previous approaches of inverse optimal control, extending them towards a more general inference of complete optimization models. These models involve constraints, partially unknown dynamics, and possible changes in \textit{modus operandi}. 

This paper \sreview \sout{follows} has \ereview the following structure: Section~\ref{sec_biology} delves into optimality principles in biology, including a historic survey of concepts and related discussions. 
Readers can approach this section with the intuitive notion of optimality as something that represents the best among all alternatives, while inverse optimality is framed as the task to inferring optimality from data.
In Section~\ref{sec_gIOC} we introduce the \sreview \sout{novel} \ereview methodology of \textit{generalized inverse optimal control}, providing a more technical exploration. Here, we also offer an introductory survey of important concepts and definitions in optimization, optimal control, \sreview and inverse reinforcement learning. \ereview 
Finally, the paper concludes with Section~\ref{sec_discussion}, where we summarize our findings and provide concluding remarks.


\section{Optimality principles in biology} \label{sec_biology}


Biology encompasses a vast array of scales, ranging from the molecular level to planetary ecosystems. On \sreview the \ereview one hand, one might be tempted to extend principles such as least action or maximization of entropy from physics and chemistry to all these biological scales. On the other hand, \textit{emergence} is a well-known phenomenon, dating back to Anderson's seminal paper \textit{More is different} from 1972 \cite{anderson1972more}. 
Presently, the emergence across scales is widely acknowledged \cite{strogatz2022fifty}. Therefore, a more nuanced examination of optimality in biology is warranted. This is particularly crucial given the processes of training and the evident evolutionary advantage associated with optimality, which serve as additional contributing factors.

We begin by presenting a brief historical survey of important milestones in the understanding of optimality in biology. Following this, we engage in a critical discussion of controversial issues providing concepts and key assumptions to address these criticisms and to advocate for the concept of \textit{generalized inverse optimal control}.  This serves as a general methodology, which will be elaborated upon in the subsequent section.

\subsection{A short history}

\subsubsection{Optimization}

One of the earliest optimality conjectures in biology was proposed by the Roman scholar Marcus Terentius Varro. Varro's conjecture aimed to explain the hexagonal cells in bees' honeycombs \cite{Mackenzie1999-kv}. Contrary to contemporary theories suggesting that bee's hexagonal lattice formation was due to their possession of six legs, Varro hypothesized that bees adopt this structure because it is the most efficient way to store honey while using the least amount of wax. Darwin provided theoretical support for this conjecture by arguing that the hexagonal design of the honeycomb is an example of adaptation that has evolved over time. According to Darwin, this design offers bees an advantage: it enables them to store honey more efficiently and thus enhances their chances of survival and reproduction, increasing the likelihood of passing their genes to the next generation \cite{Davis2004-vx}. Interestingly, the mathematical proof validating this hypothesis was only presented relatively recently \cite{Mackenzie1999-kv,Hales2001-hj}. Further examples of optimal structures in biology include tubular bones, teeth, and compound eyes \cite{McNeill_Alexander1996-df,Sutherland2005-sj}. Optimality principles have also been used to explain the \textbf{structures, movements, and behaviors of animals} \cite{McNeill_Alexander1996-df,McFarland1977-vt}. The overarching hypothesis posits that living systems have been molded by the optimizing processes of evolution \cite{Parker1990-zj}.  

Throughout history, different researchers have attempted to quantify biological systems using \textbf{physical laws} linked to optimality principles. As early as the 1920s, Alfred Lotka \cite{Lotka1922-qd} highlighted the critical role of available energy in the struggle for survival and evolution. He proposed the theory suggesting that the Darwinian concept of natural selection could be quantified as a physical law, which he termed the \textit{Law of Evolution as a Maximal Principle} \cite{Lotka1945-ds}. Lotka argued that organisms equipped with more efficient energy-capturing mechanisms gain a competitive edge. According to Lotka \cite{Lotka1922-qd}, these ideas had already been suggested by Boltzmann. Later, the principle was dubbed \textit{maximum power principle} by the systems ecologist Howard T. Odum \cite{Tilley2004-df}. This principle suggests that biological systems tend to evolve in ways that maximize their power intake or energy flux. Organisms proficient in capturing and utilizing energy resources are more likely to flourish and propagate, thereby driving the process of evolution forward.
In the 1960s, Nicolas Rashevsky, usually considered as the father of mathematical biology, introduced the optimal design hypothesis: given some prescribed biological functions, an organism possesses the most optimal possible design feasible in terms of material and energy costs \cite{Cull2007-jr}. His ideas were subsequently extended by Robert Rosen in what is  arguably the first monograph devoted to optimality principles in biology \cite{Rosen1967-nt}. Rosen was a pioneering figure who suggested that \textbf{optimal control theory} could explain the optimal dynamics of biological homeostasis and adaptation. However, he did not provide any examples of how to apply this approach nor did he provide a mathematical generalization. In any case, it is noteworthy that he proposed optimal control as a unifying tool to address various biological problems, especially considering how recent optimal control theory was at the time \cite{Sussmann1997-fu,Pesch2009-tq}. We will revisit the use of optimal control in biology later in this discussion.

\bigskip
More contemporary efforts to bridge physics and biology have considered the idea that biological systems may have evolved to optimize the gathering and representation of \textbf{information} \cite{Tkacik2009-co,Bauer2022-fh}. Recently,  studies \cite{Wong2023-rx,Sharma2023-dy} have suggested possible explanations of biological evolution with core extremal principles based on information. Wong et al. \cite{Wong2023-rx} introduce a principle termed the \textit{law of increasing functional information}. According to this principle, the functional information within a system will increase (indicating evolution) when numerous system configurations are selected for one or more functions. The level of functional information should increase in proportion to the degree of function, starting from zero for no function (or minimal function) and reaching a maximum value equivalent to the number of bits required to precisely define any system configuration that is both necessary and sufficient. Additionally, Sharma et al. \cite{Sharma2023-dy} introduced \textit{assembly theory} (AT) as an interface between physics and biology. This theory aims to explain the emergence of complexity and evolution in nature by quantifying the degree of causation required to produce a given ensemble of objects. AT emphasizes on the concept of \textit{minimum viable memory} (MVM), which refers to the minimal amount of information required for a system to self-assemble and maintain its structure.

In molecular systems biology, the use of optimality principles has been driven by the idea that molecular networks and processes are optimized for \textbf{simplicity and efficiency}. It is believed that these optimality principles serve as valuable tools for explaining and predicting biological phenomena at the molecular and cell levels. Early works exemplifying this approach can be found in references \cite{savageau1974optimal,melendez1988economy,melendez1994optimization,Heinrich1991-gv}. Subsequently, mathematical optimization has been succesfully used across a wide variety of topics, encompassing model building, optimal experimental design, metabolic engineering, and synthetic biology \cite{hatzimanikatis1996optimization,mendes1998non,torres2002pathway,Banga2008-ay,zomorrodi2012mathematical,De_Vos2013-ew,Schuster2013-tk}.

Moreover, optimality studies in systems biology use detailed models to understand  the underlying reasons why biological systems function as they do, rather than solely focusing on how they behave over time (see Chapter~6 of \cite{Heinrich1991-gv} and Chapter~9 of \cite{KlippTextBook2016} for a discussion on optimality and evolution in the context of systems biology). This approach transcends the basic mechanics (dynamics) of the system and focuses on its overall \textbf{purpose} (\textbf{function}) within the cell and its contribution to survival (fitness). The ultimate purpose of these studies is to inquire: what is the theoretical maximum benefit (fitness) that a system could achieve, and how would it be designed? If a real system closely resembles this theoretical optimum, it suggests that natural selection may have optimized it for that purpose \cite{KlippTextBook2016}. In this way, researchers are able to delve deeper into specific systems, identifying which features could potentially be adapted and what limitations (constraints) exist on those adaptations. For recent detailed studies on several aspects of optimality assumptions and constraints in the context of cell biology, refer to \cite{perez2009structure,bruggeman2023trade,scott2023shaping,EcoCellCollec_2024}.

One of the most well-known application of optimization in systems biology is probably metabolic flux balance analysis (FBA) \cite{Varma1994-zc}. FBA and its variants have shown strong agreement with experimental data across both natural and perturbed metabolic pathways \cite{segre2002analysis,ibarra2002escherichia,Shlomi2005,lewis2010omic}. This technique uses mathematical optimization to analyze the behavior of metabolic networks under the assumption of steady state. By making this assumption, FBA simplifies the system and enhances computational efficiency, eliminating the need for detailed kinetic information about the reactions in the system. Typically, FBA uses linear programming to identify network states that are optimal relative to a defined objective. This approach allows for its application to genome-scale models \cite{Price2004-ot}. However, it also means that standard FBA cannot capture the dynamics of the system or the effects of transient changes in metabolite concentrations. In cases where dynamics are important, transitioning from steady state optimization to dynamic optimization (also known as optimal control) becomes necessary.

\subsubsection{Optimal control}

\begin{figure*}[ht!!!]
  \centering 
  \includegraphics[width=10cm]{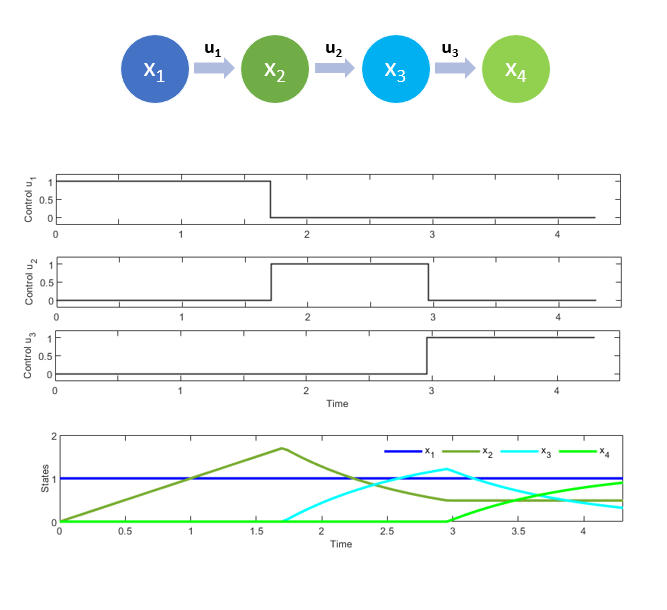} 
\vspace*{-0.5cm}
  \caption{Optimal control in a simple three-step linear metabolic pathway, where substrate $x_1$ has a buffered concentration and is converted into product $x_4$ in three steps catalyzed by three enzymes $u_i$. The upper plots show the optimal controls (concentrations $u_i(t)$) that minimize transition time (i.e., time to convert a certain amount of product). Due to a path constraint on the total amount of enzyme at any given time, the optimal enzyme profiles show a pulse-like behavior. This pattern agrees with experimental results of temporal gene expression \cite{Klipp2002-sg, Zaslaver2004-rd}. Details of the formulation are given in \cite{Tsiantis2020-nf}, based on previous works \cite{de2014global,Bartl2010-vt, Klipp2002-sg}. Variants of this problem are discussed in Section~\ref{sec_metabolic}.} 
  \label{fig:panelLinearPathway} 
\end{figure*}

 The use of optimal control theory in biology has undergone evolution, from its early acknowledgment by Rosen as a unifying framework to its application in explaining cellular phenomena and metabolic systems. For an introduction to the mathematical aspects of optimal control and its application to various biological systems, refer to \cite{lenhart2007optimal}. Optimal control theory gained momentum in the 1970s with applications in optimal decision-making in biomedical engineering, particularly in areas such as drug scheduling \cite{swan1981optimal,bellman1983mathematical}. For instance, the Norton-Simon hypothesis, which suggests that \textit{Chemotherapy success is proportional to the growth rate of proliferating cancerous cells}, led in the late 1970s to the recommendation of early, dense, high-dosage chemotherapy treatments for breast cancer. This marked a significant success for mathematical modeling \cite{Simon2006b}. Optimal control applied to designing drug regimens for disease treatment (especially cancer) remains one of the most studied problems in biomedical engineering \cite{martin1994optimal,engelhart2011optimal,pefani2013design,Moore2018,Sharp2020,Jost2020}.

Another area where optimal control theory found application as early as the 1970s is in the analysis of \textbf{animal behavior}, including reproduction, gait, and foraging. For example, McFarland analyzed the mating behavior of newts by applying Pontryagin's maximum principle \cite{McFarland1977-vt}. Moreover, he recognized the usefulness of an inverse approach, which involves identifying the unknown objective function of the assumably optimal behavior from data. The analysis of animal gaits began with basic estimations of the number of legs making contact with the ground \cite{McFarland1977-vt} and has resulted in detailed animal-specific \cite{wampler2009optimal} and human-specific \cite{Mombaur2017-ur} dynamic analysis using multi-body mechanistic models.

Optimal control theory has also found application in explaining cellular phenomena and predicting dynamics in \textbf{molecular and cell biology}. Early works considered the optimal control of bacterial growth using highly simplified models \cite{DAns1971,bley1981optimal}. In the 1980s, cybernetic models which where based on optimal control and low-dimensional dynamic models, were used to successfully predict diauxic growth in bacteria \cite{Ramkrishna1983,Kompala1984,Kompala1986,dhurjati1985cybernetic}. Cybernetic modeling of metabolism was subsequently extended in various ways, as reviewed in \cite{Ramkrishna2012}.
Other researchers adopted alternative strategies to incorporate more realistic details in optimization-based frameworks. They operated under the assumption that biological cell functions are governed by the dynamics of different biomolecular pathways, including gene expression and regulation, metabolic networks, and cell signaling pathways. In this context, optimality principles have been used to explain the structure and regulation of metabolism \cite{Heinrich1996-ni,Kremling2015}. Dynamic Flux Balance Analysis (dFBA) was proposed \cite{Mahadevan2002-ft} as a dynamic extension of FBA and can be formulated as an optimal control problem \cite{Hoffner2013-vj}.

Klipp et al. \cite{Klipp2002-sg} used optimality-based methods to explain the dynamics of simple metabolic networks, predicting temporal gene expression. Their approach involved considering simple linear networks and the hypothesis of minimum transition time. Notably, their predictions were subsequently \textbf{experimentally confirmed} \cite{Zaslaver2004-rd,chechik2008activity}, demonstrating the predictive value of optimality-based methods in systems biology. The problem considered by \cite{Klipp2002-sg} was later reformulated as an optimal control problem, leading to the study of variants \cite{Oyarzun2009-ef,Bartl2010-vt,de2014global}. Figure~\ref{fig:panelLinearPathway} illustrates an example of a simple linear metabolic pathway. These studies have stimulated a proliferation of applications of optimal control in biochemical pathways, as reviewed by \cite{Ewald2017-ws,Tsiantis2020-nf}, and in cell models incorporating protein synthesis and growth \cite{Giordano2016-cm,Ewald2017-ws,EcoCellCollec_2024}.

\begin{table*}[t] 
\begin{tabular}{lll}
\rowcolor[gray]{0.85}
\textbf{Scale} & \textbf{Description} & \textbf{References} \\ \hline
\textbf{Molecular} & Protein folding: native state of a protein has the lowest free energy & \cite{Anfinsen1973-qw,kuhlman2019advances}                                                                                        \\
&  Optimality of the genetic code                                                 & \cite{freeland2000early,itzkovitz2007genetic}                                                                                   \\
&  Optimal genome size in bacteria                                                & \cite{ranea2005microeconomic}                                                                                 \\
\hline 
\textbf{Pathway} & Optimality in the design of metabolic networks                          & \cite{savageau1974optimal,melendez1988economy,melendez1994optimization,heinrich1998modelling,noor2010central} \\
& Flux Balance Analysis (FBA) and variants predict optimal growth in bacteria     & \cite{segre2002analysis,ibarra2002escherichia,Shlomi2005,lewis2010omic}                                                  \\
&  Optimal NADH/NADPH specificities seem to maximize thermodynamic driving forces     & \cite{Bekiaris2023}                                                  \\
& Dynamic flux balance analysis explains diauxic growth in bacteria               & \cite{Mahadevan2002-ft}                                                                                       \\
& Optimal control explains temporal gene expression in simple metabolic pathways  & \cite{Klipp2002-sg,Zaslaver2004-rd,Oyarzun2009-ef,Bartl2010-vt}                                               \\
& Optimal control explains dynamics and regulation in biochemical pathways        & \cite{berkhout2012optimality,Ewald2017-ws,Tsiantis2020-nf,Kobis2022}                             \\
& Optimal design in the signalling network of bacterial chemotaxis                & \cite{kollmann2005design,celani2010bacterial}                                                                                     \\
\hline
\textbf{Cell} & Optimality theory for microbial physiology                                 & \cite{Bruggeman2020-fu}                                                                                       \\
& Optimal control explains dynamic allocation of cellular resources               & \cite{Giordano2016-cm,EcoCellCollec_2024}                                                                      \\
& Optimality of the expression levels of a protein in bacteria                    & \cite{dekel2005optimality}                                                                                    \\
& Optimal control for proteome adaptation in bacteria                             & \cite{pavlov2013optimal}                                                                                      \\        
& Optimal control explains dynamic allocation of resources in cyanobacteria                           & \cite{Reimers2017}                           

        \\
        & Optimality in mitochondrial dynamics 
        & \cite{Chustecki2024}                           

        \\

\hline
\textbf{Tissues and} & Optimal control explains the development of intestinal crypts & \cite{itzkovitz2007genetic}                                                                                  \\
\textbf{organs} & Optimization explains patterns of cell division that minimize risk of cancer   & \cite{frank2003patterns}                                                                                      \\
 & Optimality of teeth and bone structures                                        & \cite{Sutherland2005-bh,currey1985thickness}                                                                  \\
 & Optimality in the vascular system                                              & \cite{murray1926physiological,zamir1976optimality,djonov2002optimality,DuranNebreda2020}                                       \\
 & Optimization in the evolution of eyes                                           & \cite{goldsmith1990optimization,Land1992}                                                                              \\
 & Optimal design in compound eyes                                                & \cite{snyder1977acuity}                                                                                       \\
 & Free energy principle in neuroscience                                          & \cite{Friston2006,friston2010free,Friston2023}                                                                                        \\
\hline
 & Optimal behavior and life-styles in animals                                    & \cite{janetos1981imperfectly,McNeill_Alexander1996-df}                                                                              \\
\textbf{Organism}  & Optimality in the foraging behaviour of animals                       & \cite{pyke1977optimal,krebs2019foraging}                                                                          \\
 & Optimality of human gait                                                       & \cite{ackermann2010optimality,clever2018humanoid}                                                                                \\
 & Optimal control for muscoskeletal simulation                                   & \cite{Berret2018,dembia2020opensim}                                                                                      \\
& Optimality in sensorimotor control                                             & \cite{todorov2004optimality,Berret2008,Nagengast2009,Berret2016}                                                                                  \\
 & Optimality of gas exchange in plants                                           & \cite{hari1999field}                                                                                          \\
 & Optimal nitrogen distribution within a plant canopy                            & \cite{hikosaka2016optimality}                                                                                 \\
 & Whole-plant optimality predicts changes in leaf nitrogen                       & \cite{caldararu2020whole}                                                                                     \\
\hline
\textbf{Population} & Optimal sex ratio                                                    & \cite{karlin1983optimal,Smith1978-dn}                                                                         \\ 
 & Optimality in behavioral biology                                               & \cite{mcnamara2001optimality}                                                                                 \\
 & Optimal allocation of resources in a wasp colony                               & \cite{McNeill_Alexander1996-df}                                                                              \\
 & Optimal growing and breeding strategies                                        & \cite{McNeill_Alexander1996-df}                                                                              \\
\hline
\textbf{Ecosystem}  & Optimality in microbial consortia                                              & \cite{harcombe2014metabolic,zomorrodi2012optcom,theorell2022metabolic,khandelwal2013community}                \\
& Thermal optimality of ecosystem respiration                           & \cite{chen2023evidence}                                                                                       \\
 & Optimal foraging in marine ecosystem models                                    & \cite{visser2013optimal}                                                                                      \\
 & Optimality theory predicts acclimation of photosynthetic capacity in plants    & \cite{smith2020mechanisms}                                                                                    \\
 & Optimality in plant ecology                             & \cite{makela2002challenges}                                                                                 \\
 & Eco-evolutionary optimality in vegetation dynamics                             & \cite{franklin2020organizing}          
\end{tabular}

\caption{Illustrative compilation of examples of optimality principles operating at various scales within biological systems. This list is by no means exhaustive. In several cases, review papers are cited. \label{tab_optimality}}
\end{table*}

Bioprocess engineering represents another key area for the application of optimal control \cite{Smets2002,banga2005dynamic,Yegorov2018,harmand2019optimal,Yabo2022}. A recent example of optimal control applied to metabolic pathways is the microbial production of polyhydroxyalkanoates using different carbon sources \cite{duvigneau2022multiscale}. Moreover, in recent years, the fields of synthetic biology and metabolic engineering have been providing new tools and approaches for advancing bioprocess engineering. However, the application of control engineering approaches in synthetic biology to optimize, analyze, and support the design of metabolic networks has only recently gained attention and may still be in its early stages \cite{cosentino2011feedback,He2016,Steel2017,DelVecchio2018,mauri2020enhanced,gutierrez2022dynamic,Boada2022}.

In summary optimal control can be applied across different scales (in time and space) and using models of different granularity, depending on the questions being addressed, the underlying assumptions, and the available data used to validate explanations or predictions. These models can range from simplified ones that provide a broad overview of metabolism to more complex ones that aim to capture the intricacies at various molecular levels, or even at the whole cell level. The recent emergence of large-scale quantitative proteomic data, facilitating precise quantification of the actual proteomic cost with specific cellular operations, is expected to further enhance these approaches. Additionally, optimization approaches leveraging cell models at various molecular levels, including whole-cell models, are complementing these efforts \cite{Ewald2017-ws}. This \textit{multi-scale optimality} approach can be regarded as a natural extension of multi-scale modeling \cite{dada2011multi,hasenauer2015data,montagud2021systems}.
The functional scale at which optimization occurs is determined by the level at which natural selection operates \cite{Foster2011}. However, when confronted with a new problem, how do we determine the appropriate model scale and granularity? Is a multi-scale approach necessary? The study of complex systems has imparted valuable lessons \cite{goldenfeld1999simple}, encapsulated by ``\textit{Don't model bulldozers with quarks}'', i.e., choose the appropriate level of description to accurately represent the phenomena of interest. Viewed from this perspective, a multi-scale approach only makes sense if the relationships and interactions of components at different scales are really necessary to describe the observed dynamics. Once more, this will usually depend on the intended use (questions to be addressed) and the available data and prior knowledge. In addition to the whole-cell models mentioned above, another relevant example of the usefulness of a multi-scale optimality approach is eco-evolutionary optimality in vegetation dynamics \cite{franklin2020organizing}. In this context, there exists a clear necessity for analyzing the appropriate temporal and spatial scales, where organ-scale optimality (e.g., leafs) is nested within whole-plant optimality. Table~\ref{tab_optimality} provides a non-exhaustive list of examples illustrating optimality principles operating at various scales within biological systems.

\subsubsection{Reverse optimality}

Despite the general success of optimality principles, one fundamental challenge in biology lies in the fact that the objective function $\phi$ (representing the performance index to be optimized or the associated costs) is usually unknown in advance. For example, in \textbf{metabolic networks}, common candidates for objective functions include maximizing specific fluxes, minimizing transition times, minimizing intermediates, or maximizing efficiencies, among others \cite{Heinrich1991-gv}. However, it remains unclear a priori which objective function is relevant for a particular pathway. Traditionally, researchers have found these cost functions through costly trial-and-error cycles of experimentation and modeling. This involves extensive and time-consuming iterations between experimental work, data analysis, and model-based guessing of the objective function. Additionally, in some cases, such candidate functions are not well characterized (e.g., cell signaling). For the case of steady state metabolic systems, several studies have proposed an optimization-based framework to evaluate whether experimental fluxes align with candidate objective functions \cite{Burgard2003-ml,Zhao2015-zh}. However, for the broader and more intriguing task of inferring objective functions from dynamic data in molecular biology, to the best of our knowledge, only \cite{Tsiantis2018} has suggested an inverse optimal control approach, similar in spirit to the ``reverse optimization'' suggested by McFarland \cite{McFarland1977-vt} in the context of animal behavior. 

Understanding the specific optimality principle that governs subsystems within such complex interactions would represent a substantial \textbf{game-changer} in bioprocess engineering and industrial biotechnology. This knowledge is analogous to its initial application in human-centered robotics and autonomous driving. Here, the engineering goal is to mimick human behavior to achieve wider acceptance. Therefore, understanding what humans (approximately) optimize when they are moving or driving would be immensely beneficial. Similarly, this principle applies to medical studies. Hatz showed that cerebral palsy patients optimize different objective functions while walking (compared to a control group) and discussed how this knowledge could inform medical interventions \cite{hatz2014efficient}.

\subsection{Criticism and controversy}

Although optimality principles in science are generally uncontroversial, their application and interpretation have been subject to debate, especially in the field of biology.

Shoemaker \cite{Schoemaker1991-nt} examines the strengths and weaknesses of optimality as a \textbf{metaprinciple in science}. The author analyzes Fermat's principle of least time as a case study, highlighting the interplay between teleological and causal explanations. Additionally, the author examines potential biases from the flexible nature of optimality considerations, including selective search for confirming evidence and confusion between prediction and explanation. Furthermore, Shoemaker discusses the role of optimality as an epistemological organizing principle. Overall, the paper offers a critical examination of optimality as a guiding heuristic in scientific inquiry. Following this analysis, there is an interesting section with a set of open review commentaries, showing a broad spectrum of responses from various disciplines including psychology, biology, philosophy, mathematics, and economics. These commentaries exhibit considerable diversity, revealing no strong correlation between discipline and attitude towards optimality as a heuristic. However, there was a notable lack of consensus about the usefulness, role, and epistemological basis of optimality principles. In response, the author highlights the need for improved criteria to assess the utility and validity of optimality models, as well as the usefulness of comparing and generalizing across different disciplines.

In the realm of biology, optimality principles have sparked intense debate and controversy among the scientific community. While these principles effectively explained numerous features of biological systems, they have also faced criticism for their perceived \textbf{oversimplification} and failure to accommodate the intricate details of genetic and other underlying mechanisms. The limitation of optimality models in neglecting genetic information has been widely acknowledged and debated \cite{Bull2010-kb}, with some arguing that genetic variation plays a crucial role in evolutionary processes that cannot be overlooked. In their famous paper ``The Spandrels of San Marco and the Panglossian Paradigm: A Critique of the Adaptationist Programme'' Stephen Jay Gould and Richard C. Lewontin critiqued the adaptationist program in evolutionary biology \cite{Gould1979-hu}. The paper used the analogy of spandrels in architecture to challenge the adaptationist perspective, which posits that all features of organisms are necessarily adaptations molded by natural selection. Gould and Lewontin argued that numerous features in organisms are not direct adaptations but rather by-products or constraints of other evolutionary processes, akin to architectural spandrels, which are incidental features resulting from the construction of domed ceilings.

The paper by Gould and Lewontin \cite{Gould1979-hu} sparked a debate where numerous researchers supported the use of optimality theory in biology, including evolutionary biology \cite{Smith1978-dn,Parker1990-zj,Alexander2001-tr,Sutherland2005-sj}. Overall, these studies acknowledged the role of constraints, and their role in evolution, although some argued these constraints might themselves be adaptations.
Indeed, although most of these researchers viewed adaptationism as a powerful tool for explaining evolution, they recognised the need to consider constraints within a broader evolutionary framework. Furthermore, they advocated for a more nuanced understanding of how constraints, adaptations, and other factors interact in shaping evolution, with optimization theory being proposed as the most powerful framework for integrating such concepts. 

More recently, there have been suggestions for a potential reconciliation \cite{Birch2016-tp}, departing from the search of universal optimality principles justified solely by theoretical arguments. Instead, there is a shift towards a more promising approach: identifying maximization principles that apply conditionally, and subsequently demonstrating that the necessary conditions for these principles were met in the evolution of particular traits or behaviors. 

For instance, most optimality studies aiming to elucidate and predict the evolution of organisms have overlooked genetic details. This omission is a common source of criticism. Experimental adaptation of model organisms provides a new way for testing optimality models while simultaneously integrating genetics \cite{Bull2010-kb}. This approach holds as particularly useful for organisms with well-understood genetics. An effective illustration of this approach considers examining evolutionary processes through microorganisms, conducting controlled and repeatable experiments with viruses, bacteria, and yeast \cite{Elena2003}. A prominent case is the Long-Term Evolution Experiment (LTEE) with \textit{Escherichia coli} \cite{Barrick2009}, which has shown that, even in a constant environment, the mapping between genomic and adaptive evolution is complex and sometimes counterintuitive.
An obstacle to this latter approach is that the majority of bacteria have not been cultivated in laboratory settings. Nevertheless, there has been a recent increase in the sequencing and analysis of bacterial genomes, encompassing genomic research on strains and communities that have not been cultured. These additional experimental findings pave the way for new avenues of comparison within bacterial groups and between them, allowing exploration into the evolutionary forces driving molecular alterations and contributing to the observed diversity across various ecological scenarios \cite{Wernegreen2015}.

All this new experimental evidence facilitates a more nuanced, \textbf{empirically-grounded} approach, recognizing that identified optimality principles must be examined and validated based on available data and specific evolutionary circumstances. Similar arguments have been posited elsewhere \cite{Bull2010-kb,De_Vos2013-ew}, advocating for experimental adaptation of model
organisms and synthetic biology as new ways for testing optimality models integrated with genetics. These proposals are particularly compelling when combined with modern concepts from data science, such as a clear separation of data in training and validation sets. They are consistent with the method we present in Section 3, addressing the criticism that optimality in biology can not be tested.

\subsection{Clarifying concepts}

To structure the development of a novel approach to address above criticism, we introduce different approaches to optimality in biology, discuss their usage, and introduce some terminology.

\subsubsection{Approaches}
In biology, the concept of optimality can be employed in two distinct ways.
The first approach focuses on explaining the \textbf{evolutionary trajectory} of an organism over generations using an optimization framework. Natural selection serves as the optimizing force, driving the population toward traits that maximize fitness within the constraints of the environment and existing genetic variation. 
The second approach utilizes optimality to explain the current state of an organism. It focuses on explaining the organism's current status in terms of optimality or near-optimality following a lengthy evolutionary process, as studied in the first approach. This \textbf{evolutionary outcome} perspective assumes that after extensive evolutionary refinement, organisms have attained a state of near-optimality for their environment. This concept aligns with optimal adaptation, where traits are considered optimal relative to the selective pressures the organism faces.

Here, our primary interest lies in the second approach: understanding the current evolutionary outcome of an organism or biological system in terms of optimality. However, in many situations, it can be helpful to complement it with the first approach. 
It is important to highlight that although evolution is fundamental to the hypothesis that a biological subsystem functions optimally, in practical scenarios of the second approach where the aim is to analyze a system that has evolved over many generations in a stable environment, considering evolutionary effects during the observation period may not be necessary. Alternatively, when exploring the evolutionary history of organisms, applying optimality principles can provide insights into how and why certain traits or behaviors have evolved over time.

An interesting instance that illustrates the power of natural selection as an optimization process is convergent evolution. This evolutionary process occurs when unrelated species develop similar traits, resulting in similar evolutionary outcome following different evolutionary trajectories. Consider the  example of wings in insects, birds and bats \cite{Liu2024}. Despite these species having different evolutionary trajectories, they have converged on rather similar designs (flapping wings) due to the optimizing force of natural selection for flight. The final outcome, as seen in the present state of wings, exemplifies how optimality principles can be used to understand the remarkable adaptations shaped by evolution in the current state of organisms. 
Convergent evolution is also observed in the development of echolocation in bats and dolphins \cite{Liu2010}. Despite their vastly different environments and evolutionary histories, both species have developed the ability to navigate and hunt using sound waves, demonstrating the power of natural selection in driving similar adaptations in response to comparable challenges. Although most examples consider phenotypic convergence, some have been backed up by evidence of convergence at the molecular (sequence) level \cite{Liu2010,Castoe2009,Stern2013}

\subsubsection{Uses}
The incentive to use optimization theory is not to assert that everything in biology is optimal. Rather, the aim is to infer what (or if something) is optimal (or not) enabling:
\begin{itemize}
\setlength{\itemsep}{-2pt}
    \item understanding the evolutionary trajectory and the interplay between phenotypic adaptations and genomes,
    \item understanding the regulatory mechanisms of current adaptations, and how to manipulate them in applications such as bio-engineering and bio-medicine.
\end{itemize}

In other words, optimization theory in biology allows us to assess our knowledge and understanding of the evolutionary trajectory and diversity of life forms, the mechanisms within evolved biological systems, and how they will respond to new conditions \cite{Smith1978-dn,Alexander2001-tr,Sutherland2005-bh}.
 For instance, one key insight from the comparisons of bacterial genomes is that their ``lifestyle'' (as an adaptation to their environment) significantly influences their genomes \cite{Wernegreen2015}. A notable example is that long-term mutualists (endosymbionts) of insects, living in very stable environments, have the smallest genomes \cite{Wernegreen2015,Moya2008}. Studying these minimal natural genomes can be helpful to estimate the least amount of genetic components needed to construct a contemporary, free-living cellular entity, a key step to create a living cell \cite{Moya2008}, the ultimate goal of synthetic biology.

\subsubsection{Terminology}
Previous discussions and disputes regarding the application of optimality principles in biology may have stemmed from confusion surrounding various optimization terminologies, or the employment of ambiguously defined terms. For instance, the precise definitions and boundaries of concepts like ``constraints'', ``trade-offs'', ``robustness'', and ``evolvability'' have not always been consistently established, contributing to the lack of clarity in some discussions around optimality principles.
In this context, it can be helpful to discuss and clarify in more detail several key concepts, starting with \textbf{causality, constraints, and trade-offs}. Mayr introduced the concepts of proximate and ultimate causation in biology \cite{Mayr1961-qv}. Proximate causation refers to the immediate mechanisms underlying a biological trait (e.g., resource limitations). Ultimate causation refers to the evolutionary processes that shape a trait (e.g., ecological circumstances). Proximate causes operate within an organism's lifetime, while ultimate causes involve Darwinian selection across generations. These two concepts can help us understand trade-offs in biology \cite{Garland2022-tw}: ultimate trade-offs can act through proximate mechanisms and those mechanisms can evolve over time. In other words, understanding both the how and the why of a trait can provide a more complete picture of its evolution. Further, as discussed by Birch \cite{Birch2016-tp}, it is important to distinguish optimization principles that concern what happens at equilibrium (e.g., explaining a biological system after a long period of evolution in a constant environment) from those that concern the direction of change in evolving species.

It is also relevant to note that, as discussed by Alexander \cite{Alexander2001-tr}, the structure and behavior of organisms are shaped by two optimizing processes: evolution (which maximises fitness by enhancing an organism's ability to pass on its genes), and learning through trial and error. Many animals can learn behaviors that optimize food intake and mating opportunities. The purpose of applying optimization theory is not to validate evolution or learning, but to verify and enhance our understanding of these processes in shaping organisms' characteristics and behaviors.

Another important consideration is that the application of optimality principles in biology is often based on \textbf{cost-benefit analysis}, leading to \textbf{trade-offs}. These models typically involve allocation constraints related to limited resources such as energy, time, or other resources. The presence of these constraints prevents the optimization of all fitness components simultaneously, leading to trade-offs where improving one component requires sacrificing another. Understanding the relationship between trade-offs and constraints is crucial for explaining biological phenomena and making accurate predictions about the evolution of traits \cite{Garland2022-tw}. However, trade-offs in biology are complex and multifaceted, defying a single, precise definition due to their ubiquity and the relationships among different levels of organization and causality. Instead of offering a single definition, Garland et al. \cite{Garland2022-tw} describe six categories of trade-offs, spanning various biological levels of organization and encompassing both proximate and ultimate causes.

Such trade-offs can be addressed using mathematical multi-criteria optimization, where Pareto optimality is a key concept. It describes a situation where the value of one objective function cannot be improved without impairing at least one other objective function. The Pareto optimal set represents the set of solutions where there are inherent trade-offs between conflicting objectives. Therefore, multi-criteria optimization captures the best compromises between conflicting objectives and constraints, providing a systematic and quantitative framework for identifying the set of optimal solutions that balance these trade-offs. Pareto optimality has been successfully used in various areas of biology, including evolutionary, systems, and synthetic biology, see \cite{torres2002pathway,sendin2006model,sendin2009multi,shoval2012evolutionary,Warmflash2012,schuetz2012multidimensional,otero2014multicriteria,Boada2016,seoane2017multiobjetive,SeoaneSole2015,OteroMuras2017,Karamasioti2017} and references therein for further details.

Another key concept is \textbf{robustness}, usually understood as the ability of a biological system to maintain its function despite perturbations (internal and/or external) \cite{Stelling2004,Kitano2007}. While optimality in a biological context often refers to the most efficient state or process, robustness can sometimes be achieved at the expense of this efficiency. For instance, the need for robustness can drive increased complexity in biological systems, which might not appear to be the most efficient or optimal solution from a different perspective. However, this increased complexity and robustness can provide the system with the flexibility to adapt to changes and withstand various disturbances, which could be viewed as optimal in terms of survival and persistence. Therefore, we can analyze optimality in biology as trade-offs between robustness and other system properties, such as efficiency or complexity, aligning with the cost-benefit analysis considerations discussed earlier.

In line with this perspective, Chandra et al \cite{Chandra2011} review progress towards a unified theory for complex networks that can be applied to physical, biological, and artificial systems.
They outline a proposal for a unifying theory integrating methods from robust control theory and optimization. In the particular context of biological systems, the authors discuss the formalization of \textbf{tradeoffs between efficiency and robustness}. They illustrate these concepts with a case study considering glycolysis, explaining how the observed oscillations result from autocatalysis and the tradeoffs between fragility, efficiency, and complexity. They argue that nature has evolved a feedback structure that manages these tradeoffs effectively, providing adaptability to changes in supply and demand, and robustness to noisy gene expression, but at the expense of increased enzyme complexity. Their main conclusion is that, similar to engineering, complexity in biology is primarily driven by robustness.

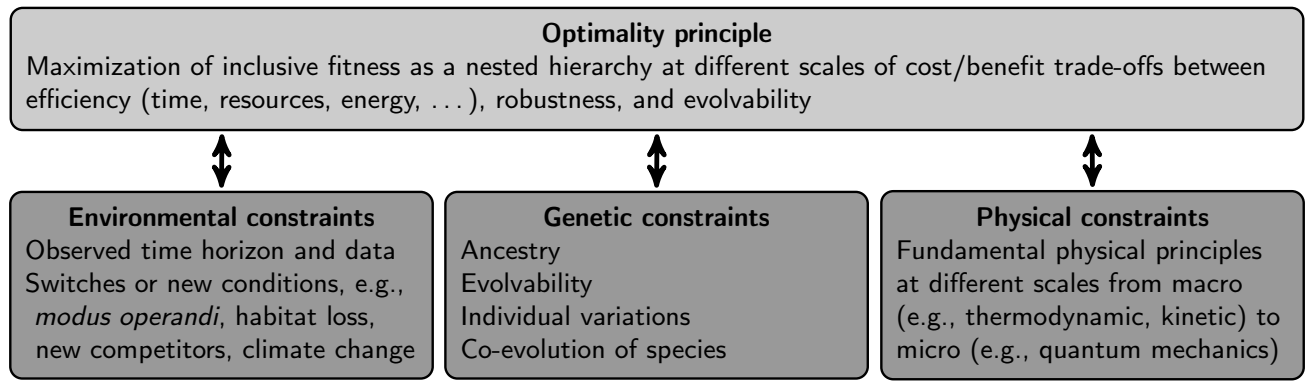
\begin{figure*}[ht]
\begin{center}
%
%
%
\resizebox{1.\textwidth}{!}{
\begin{tikzpicture}[<->,>=stealth']

\node[draw=none] (z) {};
\node[draw=none, below = 3.25cm of z] (z2) {};

\node[mybox, fill = grey2, text width=14.8cm, below = .1cm of z] (I) {\myboxcontent{Optimality principle}{Maximization of inclusive fitness as a nested hierarchy at different scales of cost/benefit trade-offs between \newline efficiency (time, resources, energy, \ldots), robustness, and evolvability}};

\node[mybox, fill = grey3, text width=4.6cm, left = 2.5cm of z2] (IIenv) {\myboxcontent{Environmental constraints}{Observed time horizon and data \newline Switches or new conditions, e.g., \newline \hspace*{0.1cm}\textit{modus operandi}, habitat loss, \newline \hspace*{0.01cm} new competitors, climate change}};

\node[mybox, fill = grey3, text width=4.6cm, left = -2.6cm of z2] (IIgen) {\myboxcontent{Genetic constraints}{Ancestry \newline Evolvability \newline Individual variations \newline Co-evolution of species}};

\node[mybox, fill = grey3, text width=4.6cm, right = 2.5cm of z2] (IIphy) {\myboxcontent{Physical constraints}{Fundamental physical principles at different scales from macro (e.g., thermodynamic, kinetic) to micro (e.g., quantum mechanics) }};

\draw[arr] ([yshift=-1pt]I.south) to ([yshift=1pt]IIgen.north);
\draw[arr] ([yshift=-1pt,xshift=-5.1cm]I.south) to ([yshift=1pt]IIenv.north);
\draw[arr] ([yshift=-1pt,xshift=5.1cm]I.south) to ([yshift=1pt]IIphy.north);

\end{tikzpicture}
}
\caption{Optimality principles in biology and their embedding into environmental, genetic, and physical constraints. Constraints (physical, genetic and environmental) and nested multi-criteria trade-offs shape the outcomes of evolutionary processes. Organisms are limited by physical laws, their genetic makeup (inherited traits), and the environment they live in (resources, predators, etc.). These constraints influence the ``optimal'' solution that can evolve. But such optimality always involves trade-offs. Balancing trade-offs is essential for the overall fitness and functionality of organisms.
} \label{fig_optimality_overview}
\end{center}
\end{figure*}

Khammash \cite{khammash2016} provides an insightful analysis of biological robustness from an engineering perspective, using electronic amplifiers and gene expression circuits as illustrative examples. These systems share remarkable similarities, with negative feedback serving as the main strategy to achieve robustness. However, he also observes that optimality, both in artificial and in biological complex systems, will often be sensitive to specific perturbations, suggesting universal trade-offs between robustness and fragility. In the same spirit, Carlson and Doyle \cite{Carlson2002} argue that this robust yet fragile nature is not an accident of evolution but a fundamental aspect of complexity.

The analysis of these trade-offs in molecular systems biology can offer a deeper understanding of the \textbf{design principles} of biological organization and their regulation \cite{Wall2004,Adiwijaya2006,Savageau2009,Higuera2012,Szekely2013,OteroMuras2016,Adler2017,Frank2019,Cao2020}. For instance, El-Samad et al. \cite{ElSamad2005} use the heat shock response in bacteria as an example, extracting design motifs and justifying them in terms of performance objectives and their trade-offs. Their analysis uses a modular decomposition that parallels traditional engineering control architectures. Such trade-offs resemble those usually considered in engineering, where typical control designs aim to simultaneous minimizing two competing objectives: deviation from some desired operation and the necessary control effort.

Finally, another intriguing aspect of biological robustness is its relationship with \textbf{evolvability} \cite{Wagner2007,Masel2010}. Evolvability refers to the capacity of a biological system to generate adaptive genetic diversity and evolve through natural selection. Wagner \cite{Wagner2007} emphasizes that evolvability is not just about the likelihood of immediate change, but also the potential for future evolutionary modification.
A system with high evolvability has a greater capacity to generate new and beneficial variations through mutations.
Robustness and evolvability are two complementary properties of biological systems essential for their survival and adaptation to changing environments. Wagner \cite{Wagner2007} suggests that robustness is less about organisms having plenty of spare redundant parts and more about the fact that mutations can alter organisms in ways that do not significantly affect their fitness. He also notes that robustness only matters for one feature: fitness, understood as the ability for survival and reproduction.
Crucially, a system that is robust and evolvable may not be the most efficient (i.e., optimal) in the short term with respect to specific features (e.g., metabolic efficiency), but it could be considered optimal in terms of long-term survival and adaptability \cite{Whitacre2010}. 

In summary, robustness, evolvability, and efficiency are intertwined in biological systems, with constraints, trade-offs, and balances between them shaping the evolution and functioning of these systems. A simplified diagram representing these interconnections is shown in Figure~\ref{fig_optimality_overview}.

\subsection{Key assumptions}

We now focus on the case of molecular and cell biology, assuming that due to evolutionary selection processes dynamic behavior may be optimal. Infererring what exactly constitutes optimality would not only help in acquiring a deeper understanding of the considered process (scientific insight) but also enable the utilization of the optimality principle in forward optimal control. 

However, as previously mentioned, this approach has faced considerable criticisms. For example, as discussed by Maynard Smith \cite{Smith1978-dn}, Lewontin argued that simply invoking an optimality principle without a clear way to define and quantify what is being optimized can result in a weak scientific explanation. Adding more constraints after the fact (the "ad hoc secondary problem") to justify a particular outcome can weaken the robustness of the model. 

These and related criticisms motivate the development of new methods aimed at inferring optimality principles directly from data. While the call for a reverse optimality approach was made several decades ago \cite{McFarland1977-vt,Smith1978-dn}, a comprehensive methodology of this nature is still lacking. We believe that the time is ripe for the development of data-driven approaches capable of analyzing extensive biological dynamic datasets to uncover the fundamental principles dictating these dynamics, while integrating constraints from the outset. In other words, we claim the necessity for a method for unbiased identification of optimality principles from dynamic (time-series) data and prior knowledge. This approach would ensure that these optimality principles are not merely speculative but grounded in empirical evidence and quantifiable constraints.
To ensure that such a method addresses most of the critiques outlined above, it would be beneficial to consider several assumptions and clarifications.

First, we assume that the optimality principle can be mathematically represented as a \textbf{constrained multicriteria dynamic optimization} problem. This formulation yields Pareto optimal solutions, providing a natural means of managing the interplay between \textbf{constraints and trade-offs}. The selection of appropriate elementary objective functions is problem-dependent and should prioritize those with \textbf{strong selection pressure}, defined as the functions expected to have the most significant impact on fitness. We are also assuming that dynamics play a key role in the optimality principle. While network structure and steady state data can define key aspects of function, robustness, and regulation in certain cases (e.g., metabolic pathways \cite{Stelling2002}), it is generally challenging to determine \textit{a priori} if this alone is sufficient to infer optimality. Therefore, we posit that a kinetic mechanistic description is necessary.

Second, \textbf{active constraints} can significantly influence the optimization process: if a constraint is binding (i.e., it is satisfied as an equality at the optimal solution), it can create a trade-off. Relaxing the constraint may enable improvement in one objective without compromising another, thereby altering the trade-off. Often these constraints are unknown \textit{a priori} and must be inferred from the available data.

Third, the choice of \textbf{time horizon} is crucial, and the concepts of proximate and ultimate causation, introduced by Mayr \cite{Mayr1961-qv}, are helpful in addressing proximate causes (relevant during an organism's lifetime) and ultimate causes (relevant during Darwinian selection across multiple generations). It is important to note that evolution is constrained by ancestry, meaning that previous evolutionary history plays a significant role. In many problems, the observation time horizon is typically rather short (e.g., in the order of the organisms's lifetime), thus aspects like evolvability can be ignored. In other cases, the observed time horizon may encompass changes (\textbf{switches}) in the underlying optimality principle. These switches can be triggered by external environmental changes or by the achievement of intermediate milestones.

An additional aspect involves situations of biological co-evolution, which can result in the so-called Red Queen effect \cite{Liow_2011-st}, i.e., evolutionary arms races between species, where they must continuously evolve and adapt to survive in a world where other competing species are also evolving. We believe that such situations could be described by a further generalization of our method in the form of generalized inverse dynamic games. However, addressing these problems is beyond the scope of this paper.

\section{Generalized Inverse Optimal Control} \label{sec_gIOC}

As outlined in the previous section, there is sginificant potential for a systematic approach to infer constrained optimality principles from observations. We are interested in formulating this task as an optimization problem and in numerical methods that allow exactly this, independently of a particular application domain. Suppose we are given observations $\eta$. We then define the generalized inverse optimal control (gIOC) problem as
\begin{equation} \label{eq_generalIOC}
\min_{\Y, \phi, y^*} \; D(y^*, \eta) \; \text{subject to} \; y^* = \arg \min_{y \in \Y} \phi(y)
\end{equation}
Giving credit to its name, this is a very general yet \textbf{abstract definition} that will be specified in the next section. Problem~\eqref{eq_generalIOC} reveals the typical bi-level structure of gIOC. On an outer level, a data fit (regression) between observed data $\eta$ and variables $y^*$ is to be minimized, using an appropriately defined distance function $D(\cdot)$. 
The degrees of freedom are the objective function $\phi$ and constraint functions specifying the feasible set $\Y$. Currently, we leave it open how these choices can be modeled in practice and how to include a priori domain knowledge and restrictions.
As a special feature, the variables $y^*$ must be optimal (the argument $\arg$ that minimizes \eqref{eq_generalIOC}) for an (a priori unknown) inner level optimization problem that is specified using $\phi$ and $\Y$ from the outer level.
The general definition \eqref{eq_generalIOC} allows to incorporate the assumptions made in the previous section, such as the inclusion of prior knowledge in the problem formulation. 

In analogy to inverse simulation, which seeks the origin of an observed behavior of a dynamic system in contrast to a forward simulation, compare Section~\ref{sec_introduction}, in gIOC one is interested in inferring optimality principles, constraints, and associated dynamic behavior from observations. Generalizing existing inverse optimal control (IOC) approaches, in gIOC, the aim is to identify \textit{all} components of an underlying optimal control problem, not only the objective function. We are thus interested in the inverse question of optimal control: how to infer a priori unknown optimality principles (or validate hypotheses concerning optimality), constraints, and partially unknown dynamics from data. In particular, we aim to develop a systematic approach to infer all symbolic and numerical unknowns from time series data. In this sense, the new class of gIOC is a \textbf{superset} of IOC and inverse reinforcement learning (IRL), but also of model identification problems \cite{camps2023discovering}.

For a better intuition of the properties of the problem class, we shall discuss its components one by one and examine special cases that are relevant in biology in Section~\ref{sec_modeling}.
We will illustrate the concept by providing examples in Section~\ref{sec_examples}.
We discuss numerical methods for solving \eqref{eq_generalIOC} and review \sreview the state-of-the-art in IOC and IRL \ereview in Section~\ref{sec_methods}. \sreview We provide links to valuable previous work especially in the robotics and model predictive control communities, which may serve as a good basis for future gIOC methods for applications in biology. \ereview

\subsection{Modeling of gIOC problems}\label{sec_modeling}

We discuss problem~\eqref{eq_generalIOC} and give specific settings concerning $y, \phi$, and $\Y$ that are most relevant for dynamic processes in molecular systems biology.

\subsubsection{Dynamic systems and variables $y$} \label{sec_dynamics}

We start with a closer look at the variables $y$ in \eqref{eq_generalIOC}.
In mathematical modeling, we usually distinguish between state (dependent) and control (independent) variables. State variables, denoted here by $x$, represent the internal state (past, current, and future condition) of the system. Control variables, denoted by $u$ and also known as input variables, have slightly different meanings depending on the application. In engineering and design, controls are the variables that can be manipulated or controlled to achieve a desired output or response from the system. In the natural sciences, they are the external forces or signals that affect the system’s behavior (i.e., as described by its states). 
The mathematical model links $x$ and $u$, allowing to calculate the dependent variables $x$ for any choice of $u$. There are many different ways to model a dynamic system, all of which can be combined with problem~\eqref{eq_generalIOC}, such as Markov chains, physics-informed neural networks \cite{raissi2019physics}, or partial stochastic delay differential-algebraic equations. Particularly relevant for molecular systems biology are \textbf{ordinary differential equations} (ODE), which we will focus on in the following for a clearer representation, but without loss of generality.
Therefore, the general variable $y$ in \eqref{eq_generalIOC} will be specified to variables $(x,u)$.
Often, the mechanics of a dynamic system can be captured with an initial value problem of the form
\begin{equation} \label{eq_ODE}
\dot{x}(t) = f(x(t), u(t)) \text{ for } t \in [0,T], \quad x(0) = x_0
\end{equation}
Initial values $x_0 \in \R^{n_x}$ for the states $x(\cdot)$ along with control trajectories, i.e., values $u(t)$ for all times $t \in [0,T]$, and assuming Lipschitz continuity of the function $f(\cdot)$, are sufficient to completely determine the future behavior of the system, i.e., to calculate all values $x(t)$ for $t \in [0,T]$. 

As an example, consider the growth of an annual plant as discussed by Maynard Smith \cite{Smith1978-dn}: the rate at which it can accumulate resources depends on its size. The resources can be partially allocated for further growth, and the rest to generate seeds. In this example, the state is one-dimensional and given by the plant size as a function of time, $x(t)$. The one-dimensional control $u(t)$ represents the fraction of incoming resources allocated to seeds at time $t$. Biological domain knowledge is needed to formulate the function $f(\cdot)$ that describes how the growth rate of the plant size depends on plant size and resources allocation. The model can then be written in the form of \eqref{eq_ODE} as a differential equation where the time derivative of $x(t)$ depends on a function $f(\cdot)$ of $x(t)$ and $u(t)$, on the initial values $x_0$ (size at time zero), and on season duration $T$. When the independent variables are chosen, lower and upper bounds should be considered, e.g., 
\begin{equation} \label{eq_ubounds}
0 \le u(t) \le 1.
\end{equation}
Relating to problem~\eqref{eq_generalIOC}, we would thus work with $y=(x,u)$ and appropriately chosen function spaces $\mathcal{X}$ and $\mathcal{U}$ to define the feasible set
\begin{equation} \label{eq_feasible}
\Y := \left\{(x,u) \in \mathcal{X} \times \mathcal{U}: (\ref{eq_ODE}-\ref{eq_ubounds}) \forall \; t \in [0,T] \right\} 
\end{equation}
which allows us to calculate $x$ from $u, x_0$, and $T$.

Modeling is by no means unique, consider the different car models compared in Figures~\ref{fig_carSimple} and \ref{fig_carAdvanced}. 
If a basic or a more refined model is appropriate depends on the questions we want to address. But it is important to note that both models are simplified representations of reality, or, in the words of George Box, ``all models are wrong, but some are useful''. Similarly, we argue that the mathematical formulation of an optimality principle is a simplified representation of a hypothesis so, in that sense, \textbf{``all optimality principles are wrong, but some are useful''}. 
In biology, the scale (or level of description) of the model can be also dictated by the available measurements. Consider the study by Klipp et al. \cite{Klipp2002-sg} to predict the dynamics in a metabolic pathway assuming optimal function under a constraint limiting total enzyme amount. The authors used a simplified dynamic model of a biochemical pathway, where the controls were the concentration of enzymes. The computed optimal controls (enzyme concentrations) were then compared qualitatively with available experimental data (gene expression profiles, which should correlate well with the enzyme concentrations). Subsequently, other authors \cite{Bartl2013-sr,Waldherr2015-rk} considered more refined models, incorporating gene expression dynamics, to obtain optimal gene expression profiles directly.

In practice, sometimes the mechanistic model underlying the dynamics, e.g., equation~\eqref{eq_ODE}, is not fully known. The equations may depend on particular \textit{model parameters} $p \in \R^{n_p}$ that enter the right hand side,
\begin{equation} \label{eq_ODEp}
\dot{x}(t) = f(x(t), u(t), p) \text{ for } t \in [0,T], \quad x(0) = x_0(p)
\end{equation}
and that allow to modify values such as reaction rates, masses, coefficients of Michaelis-Menten kinetics, initial concentrations or sizes, or similar.
The underlying assumption is, though, that a priori biological knowledge (how do states $x$ and controls $u$ relate to one another independent of the specific values of $p$) can be included in the function $f(\cdot)$.
If this is not the case, there are various options for learning the relationships from data. For example, parts of the function $f(\cdot)$ may consist of neural networks (referred to as universal differential equation \cite{rackauckas2020universal}) or algebraic functions can be learned via \textit{symbolic} regression \cite{camps2023discovering}. Here, the function $f$ (or parts of it) can be an arbitrary mathematical function composed from basis functions such as multiplication, exponential, and addition of $x$ and $u$. 
Although different algorithmic approaches are necessary as discussed in Section~\ref{sec_methods}, all of these different modeling approaches can be conveniently represented via formulation \eqref{eq_ODEp} with a general function $f(\cdot)$ depending on a parameter vector $p$. From a modeling perspective, we shall thus not further discuss the nuances of numerical regression, hybrid systems, and symbolic regression.

\subsubsection{The objective function $\phi$} \label{sec_objective}

The objective function $\phi$ can be either a functional or a function, mapping the variables $y$ to a real value $\phi(y)$. A typical formulation for optimal control problems is of so-called Bolza-type,
\begin{equation} \label{eq_objective}
\phi(x,u) := \phi_E( x(T) ) + \int_0^T \phi_L ( x(\tau), u(\tau) ) \; \mathrm{d}\tau
\end{equation}
consisting of a Mayer term $\phi_E( x(T) )$, e.g., the size of a plant at the end of the considered time horizon, and a Lagrange term $\int_0^T \phi_L ( x(\tau), u(\tau) ) \; \mathrm{d}\tau$, e.g., the overall allocation of resources until time $T$.
A point $y^*$ is considered a \textit{local optimum} if $\phi(y^*) \le \phi(y)$ in an environment of $y^*$ and \textit{global optimum} if $\phi(y^*) \le \phi(y)$ for all feasible $y \in \Y$. Finding global optima is significantly more challenging, and in difficult optimization problems one often settles with local minima. However, \textit{convex optimization problems}, for which both the objective function and the feasible set $\Y$ are convex, are an exception. Here, a local optimum automatically qualifies as a global one.
Assuming that optimal control problems can be solved locally or even globally, one can consider an optimization problem to test a hypothesis. For example, in the case of optimal plant growth from above one could ask, for a given starting size and length of season, what is the optimal $u(t)$ (allocation of resources) that maximizes the total number of produced seeds? Comparing the solution of this optimal control problem with real data, we can test the hypothesis that the plant has evolved to maximize the production of seeds.

For particular applications, it is interesting to consider multiple objectives $\phi_i$. Often, there is a trade-off, e.g., between time and energy consumption to achieve a certain goal. 
A point $y^* \in \Y$ is said to be \textit{not dominated} if there is no other $y \in \Y$ exists such that $\phi_i(y) \le \phi_i(y^*)$ for all $i$, i.e., which is better with respect to all objectives. The set of all non-dominated points is called the \textbf{Pareto front}. It is of particular interest for a posteriori decision-making as it allows balancing the advantages and disadvantages of the different objective functions $\phi_i$.

As discussed earlier, especially in biology the objective function is often not known a priori. Therefore, in \eqref{eq_generalIOC} we optimize over $\phi$ to closely match observed data. Assume that $n_\phi$ candidate objective functions $\phi_i$ are known in advance and the purpose of investigating \eqref{eq_generalIOC} is to find the point on the Pareto front \cite{torres2002pathway,sendin2009multi,shoval2012evolutionary,schuetz2012multidimensional,otero2014multicriteria,seoane2017multiobjetive,OteroMuras2017} corresponding to the best data fit, e.g., via $\phi(y) = \sum_{i=1}^{n_\phi} w_i \phi_i(y)$ with weights $w_i$ specifying the contribution of candidate objective $\phi_i$ to the objective $\phi$. Note that more efficient methods for calculating Pareto fronts exist, see, e.g., \cite{Logist2010} for further references and additional consideration of integer controls.
Returning to the example of metabolic networks, common candidates for objective functions include maximizing specific fluxes, minimizing transition times, minimizing intermediates, or maximizing efficiencies \cite{Heinrich1991-gv}. Specific numerical values of $w_i$, normalized to be between $0$ and $1$ and sum up to $1$, result in a specific compromise between these candidates. 

Among all possible objective function candidates, two stand out as particularly relevant but also as special cases concerning modeling. The first one is the \textbf{minimization of time}. 
From a mathematical point of view, treating $T$ as an optimization variable (formally, this can be easily achieved via a model parameter $p_i$ multiplied to the right-hand side of the ODE) implies another, problem-dependent modeling choice: how shall observed values $\eta_i$ at times $t > T$ be treated? One can either discard them, compare them to $h(y^*(T))$, simulate $y$ for times $t > T$ with some assumption on the applied controls, or force $T$ to be smaller than the maximum observation time.
The second one is the \textbf{maximization of robustness}. As discussed above, robustness refers to the ability of a biological system to maintain its function despite perturbations \cite{Stelling2004,Kitano2007}. In robust optimization and optimal control different approaches have been proposed, e.g., \cite{ben2009robust,gorissen2015practical,bertsimas2018data}. Important concepts include those of random and targeted attacks. Two detailed case studies of how parametric uncertainty can be used in biological networks are described in \cite{nimmegeers2016dynamic}. The authors compare linearization, sigma points, and polynomial chaos approaches. In general, all of them can be used within our gIOC approach by extending the mathematical model with additional states. These correspond either to variational differential equations (i.e., sensitivities of the states with respect to parameters) or to sample points, allowing an estimation of higher order moments of the underlying distributions. These higher-order moments can then be used to model candidate functions $\phi_i$ or candidate constraint functions $g_i$ that represent aspects of robustness.
In summary, robustness can be considered as a special case of function candidates, albeit at the expense of additional modeling effort and higher computational costs to solve problem \eqref{eq_gIOC}.
If objective function candidates are not known, this can again be addressed via formulations including neural networks or via symbolic regression. In analogy to \eqref{eq_ODEp}, this case is formally included via a dependence of the functions $\phi_i$ on the model parameter vector $p$. Whether all candidates are known or not is application-dependent. While the general formulation \eqref{eq_generalIOC} covers both scenarios, different algorithms and special care considering issues of identifiability are necessary.

\subsubsection{Constraints and the feasible set $\Y$} \label{sec_constraints}

The feasible set $\Y$ is an important feature of the inner optimization problem in \eqref{eq_generalIOC}.
There might be relevant restrictions that influence the optimal solution $y^*$.
If, for example, enzyme activity were not restricted by any physiological bounds, then metabolism could be arbitrarily fast. Hence, such restrictions need to be considered in inverse optimization approaches. In analogy to inferring objective functions, the task of gIOC is to identify such a priori unknown constraints from data.

We can distinguish three types of restrictions possibly entering the definition of $\Y$.
The first type of restrictions is related to the modeling of dependent and independent variables in the first place, as discussed above with the example of \eqref{eq_feasible}. A second type specifies what kind of values the variables $y$ may take (mathematically: to which space belongs $y$?). For example, in some applications, there may be combinatorial restrictions on the controls $u$, such as a finite number of gaits to choose from or yes-or-no inhibition of sub-processes. Combinatorial restrictions lead to the field of mixed-integer optimal control \cite{Zeile2021-cu} and require different algorithmic approaches, compare Section~\ref{sec_methods}.

While the first two types are typically addressed by a modeling expert, the third type is something that needs to be considered in the inverse optimization problem. It comprises vectors of general inequality constraints $g(y) \ge 0$. They allow to formulate lower and upper bounds on variables, but also more general relationships. 
Of particular importance is whether an inequality constraint $g_i(y) \ge 0$ is \textbf{active} in an optimal point, i.e., $g_i(y^*) = 0$, or \textit{inactive}, i.e., $g_i(y^*) > 0$. Only active constraints have a local influence on the optimum (i.e., without this constraint the optimal point $y^*$ could be different). In a practical setting, if it is unclear whether the data $\eta$ was produced by an optimal, but constrained process, one could simply solve two problems -- one with the constraint $g_i(y) \ge 0$ on the inner level, and one without. The better match to the data $\eta$ is then the more likely explanation.
However, the number of possible combinations grows quickly if there are several candidate constraint functions; therefore, we model the inclusion of constraints again as a degree of freedom for the optimization on the outer level of \eqref{eq_generalIOC}.
Assume that $n_g$ constraint function candidates $g_i(\cdot)$ are known, but it is unknown and of interest, if they were active and played a role in the observed process. One possibility to model them is via binary weight variables $w_i \in \{0, 1\}$ as
\begin{equation} \label{eq_constraints}
0 \le w_i g_i(y) \quad \forall i \in [n_g] = \{1, \ldots, n_g\}.
\end{equation}
Note that the choice of $w_i = 1$ would enforce this constraint and thus increase the (irrelevant) objective function value $\phi(y^*)$ of the inner problem, but might reduce the (relevant) data fit objective function value $D(y^*,\eta)$ of the outer problem. 
This modeling strategy is by no means unique. For example, for a real variable $w_i$ using $g_i(y) \ge w_i$ also results in an active or non-active constraint in the inner optimization problem, depending on the value of $w_i$.

\subsubsection{Interpretation of results} \label{sec_interpretationresults}

Assuming that problem \eqref{eq_generalIOC} can be solved numerically, one obtains optimal variables $y^*$ on the inner level and an optimal optimality principle $\phi^*, \mathcal{Y}^*$ on the outer level. The interpretation of these results is in spirit similar to that of other model identification or deep learning tasks. It is important to follow best practices in defining and using training, test, and validation sets as well as methods for evaluating the \textbf{significance of results} \cite{Goodfellow2016,camps2023discovering}.

The distance function $D(y^*,\eta)$ in \eqref{eq_generalIOC} can take several forms as it reflects statistical assumptions on the measurement errors. The particular choice has an influence on numerical methods as well. The most popular choice is maximum likelihood estimation, assuming independent and normally distributed measurement errors, resulting in the Euclidean norm $\| h(y^*) - \eta \|_2^2$ as the objective function on the outer level, where $h$ is an output function mapping the differential states and controls to the observables. Often, the metric distance function is extended by a regularization term $R(y)$, allowing for specifying a priori knowledge about the distribution of the unknown variables and reducing the risk of overfitting.

In the interpretation of the results, a second step to evaluate a prediction based on the identified optimization model with data from the validation set is necessary. Depending on the statistical significance of this test, there are two possible outcomes. First, no optimality is apparent. This can be due to lack of data, insufficient or erroneous modeling, or simply because there was no dominant optimality principle at work in the first place. Second, an optimality principle has been significantly identified and can be used in follow-up steps for analysis and/or usage in forward optimal control.

\begin{figure*}[ht!!!]
\centering
  \begin{subfigure}{0.43\linewidth}
  \includegraphics[width=\linewidth]{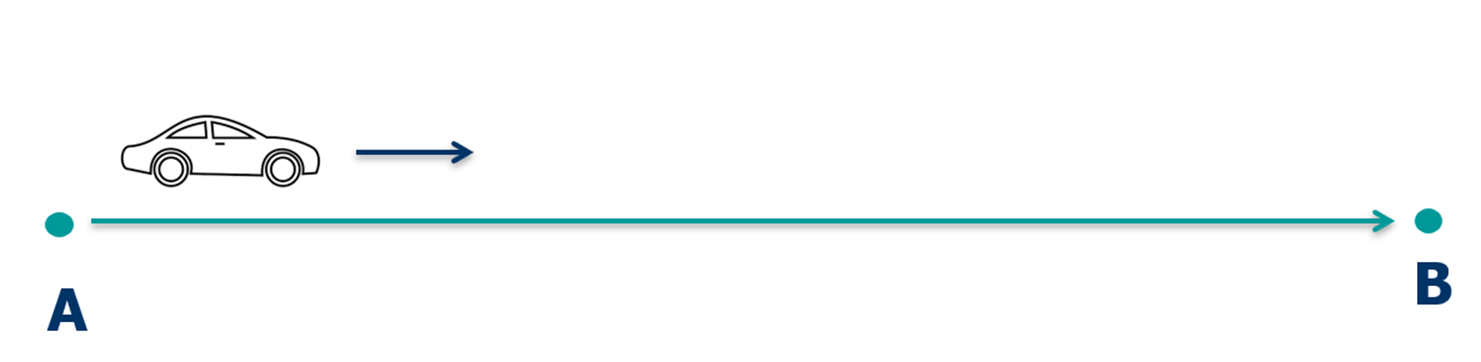}
  \caption{Task: drive in minimum time from $A$ to $B$ and stop} \label{fig:carSimpleA}
  \end{subfigure}
  \begin{subfigure}{0.53\linewidth}
\centering
  \includegraphics[width=0.7\linewidth]{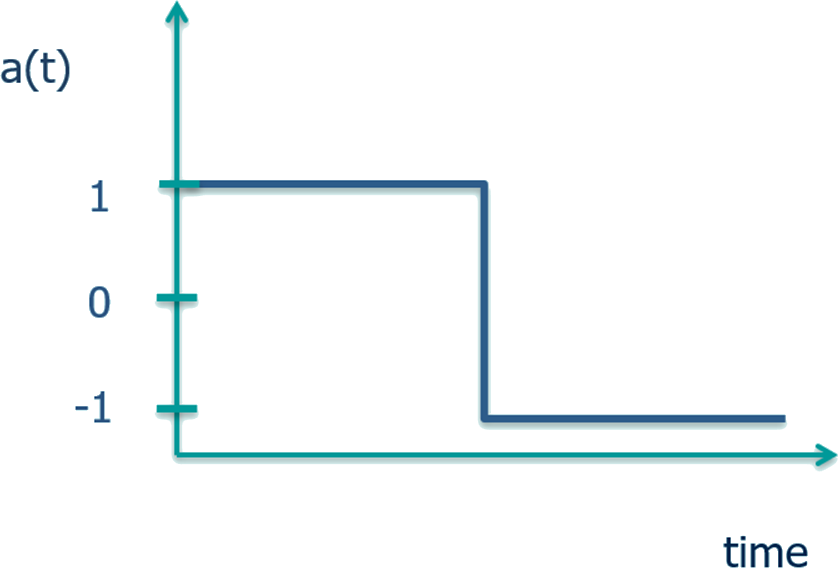}
\caption{Optimal control: maximum acceleration and maximum braking} \label{fig:carSimpleB}
  \end{subfigure}
\caption{Optimal control using a minimalistic car model in 1D. The car is at rest ($v=0$) at $t=0$ at position $x=A$. How would you accelerate/brake the car to get from $A$ to $B$ in minimum time and stop at $B$? The control $a(t)$ is bounded by $-1 \le a(t) \le 1$. The optimal control is of bang-bang type, i.e., the maximum acceleration until the midpoint is followed by maximal braking.
}
\label{fig_carSimple}
\end{figure*}

\subsubsection{Further considerations} \label{sec_further}

Optimality principles in biology seem to be nested in a hierarchical way, as discussed above. Contributing to the main goal of maximizing fitness, biological systems may have \textit{modi operandi} that are optimal for a specific function or task and the considered scale and environmental setting. Let us assume the simplified situation of a lion who knows at least three different kinds of behavior: relaxing (energy consumption minimized), sneaking up on prey (observability minimized), and hunting (time-to-target minimized). Obviously, all of them contribute to the main goal of maximizing fitness. Yet, determining the exact objective function from data may be more difficult if several \textbf{switching} objectives were used in the observed time horizon.
Such switches can be further classified.
They may either be due to external circumstances (for example, the predator suddenly observes a prey and switches from relaxing to foraging mode) or because several consecutive stages like sneaking and hunting are optimized together when foraging.
Mathematically, this difference is important. In the first case, two different optimal policies are concatenated, while in the second case the optimal policy considers all stages in one go. The difference will be illustrated in Section~\ref{sec_examples}.
Additionally, the current situation (modeled via parameters $p$) has an influence. A resting lion will react differently to a prey depending on if it is hungry or not. But in both cases, the overall cost function (fitness towards survival and reproduction) is the same.

In most situations where optimality principles are used to explain (or predict) dynamics, it is sufficient to consider the open loop optimal control problem, where in addition to the cost function we obtain the optimal control policies as a function of time, $u(t)$. For example, in the case of the simple linear metabolic pathway considered by Klipp et al. \cite{Klipp2002-sg}, we can obtain the time-dependent enzyme profiles and the optimality principle consistent with the data by using inverse optimal control \cite{Tsiantis2018}. This can be satisfactory if we simply want to explain or predict the observed dynamic behavior. 
However, if we want to obtain more information about the involved \textbf{regulatory mechanisms}, it should be noted that in reality, the enzyme concentrations depend on the expression of genes, which are themselves influenced (activated or inhibited) by the concentrations of other metabolites and enzymes, or even the expression of other genes. Thus, one might want to infer the regulatory network compatible with the observed behavior. This would be particularly useful in biotechnological and biomedical applications where we seek to change such regulations using genetic engineering. Although reduced versions of this problem have been considered, assuming certain topologies and kinetics for the feedback regulation in simplified networks \cite{De_Hijas-Liste2015-fl}, the more general problem remains, to the best of our knowledge, an open question. Here we propose gIOC as a general framework that includes these situations as a closed-loop inverse optimal control formulation.

\begin{figure*}[ht!!!]
\centering
  \begin{subfigure}{0.53\linewidth}
  \includegraphics[width=\linewidth]{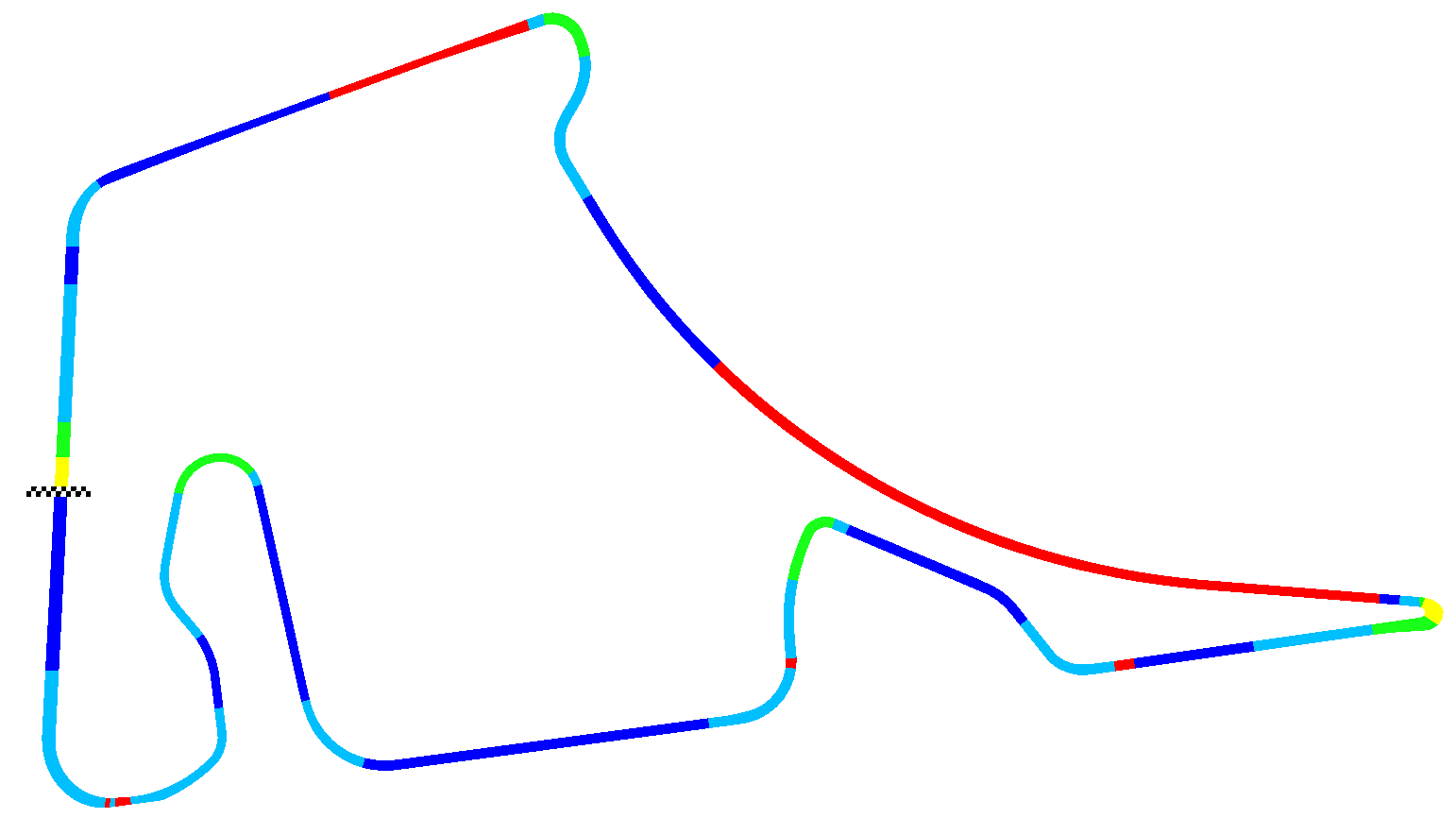}
  \caption{The Hockenheim ring and time-optimal gear choices.} \label{fig:carAdvancedA}
  \end{subfigure}
  \begin{subfigure}{0.43\linewidth}
  \includegraphics[width=\linewidth]{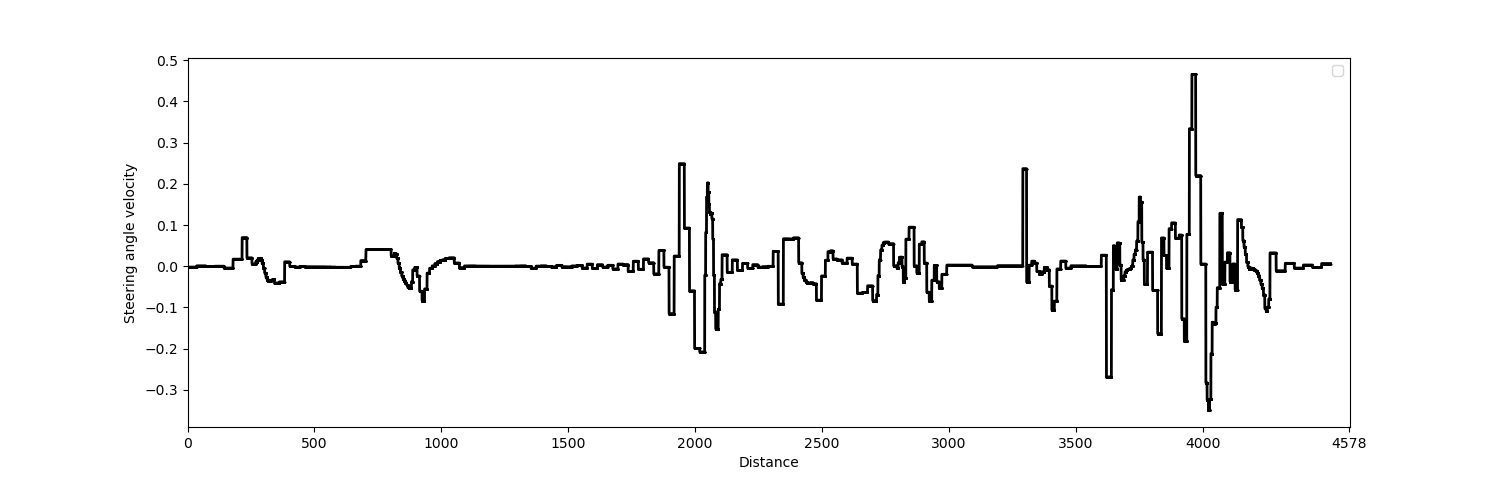}

  \includegraphics[width=\linewidth]{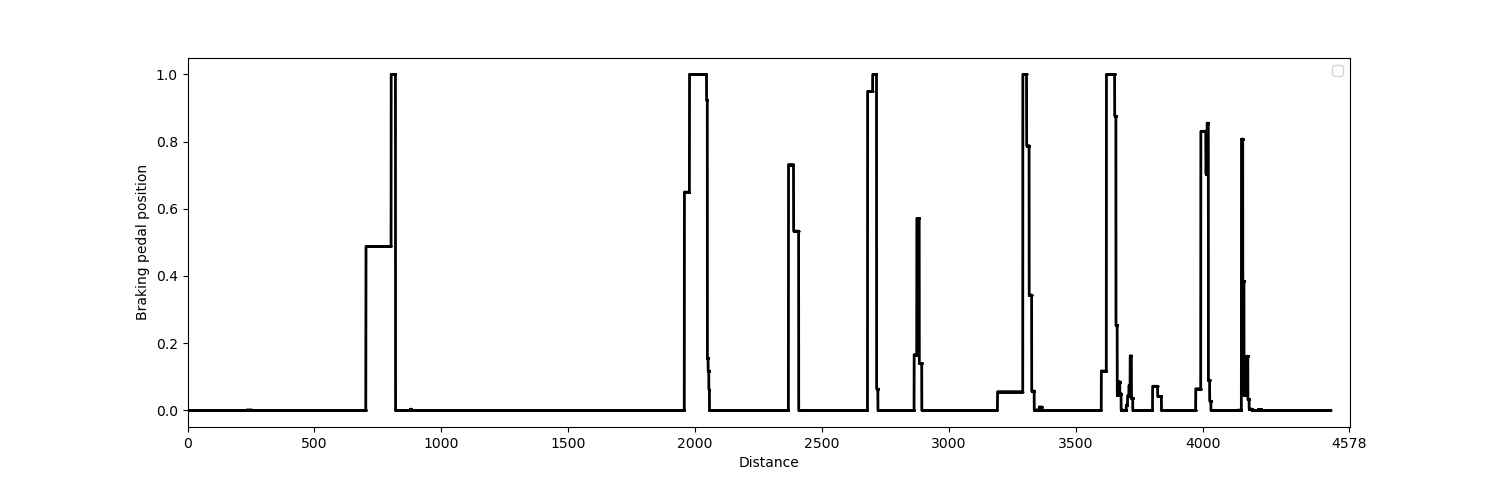}

  \includegraphics[width=\linewidth]{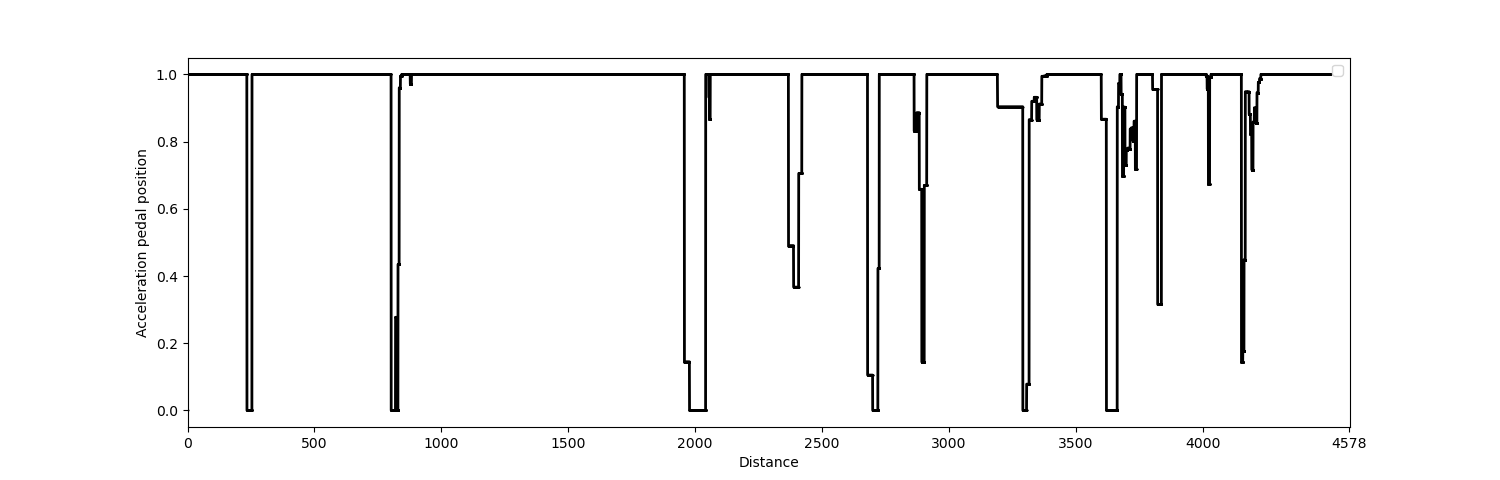}

\caption{Time-optimal additional control functions.} \label{fig:carAdvancedB}
  \end{subfigure}
\caption{Compared to Figure~\ref{fig_carSimple}, a more detailed model of a racing car (Porsche CS) on the Hockenheim ring as studied in \cite{Kehrle2011} is considered. Shown are the topology of the race track and optimal control functions calculated in \cite{Kehrle2011} that involve \eqref{fig:carAdvancedA} gear choices and \eqref{fig:carAdvancedB} steering angle velocity, braking pedal position, and acceleration pedal position. One observes that the optimality principle (here: minimum time driving) encapsulates the regulation of the system (here: the control actions of the driver) and that path constraints have an impact on the optimal controls in comparison to the simplified setting in Figure~\ref{fig_carSimple}.
}
\label{fig_carAdvanced}
\end{figure*}

\begin{figure*}[ht]
\begin{center}
\resizebox{.8\textwidth}{!}{
\begin{tikzpicture}[->,>=stealth']

\node[draw=none] (z) {};

\node[mybox, fill = grey1, text width=7.cm, left = .1cm of z] (Ileft) {\myboxcontent{Input: prior modeling knowledge}{Candidates for dynamics, objective, constraints}};
\node[mybox, fill = grey1, text width=7.cm, right = .1cm of z] (Iright) {\myboxcontent{Input: observed data $\eta$}{representable as a function of states and controls}};

\node[mybox, fill = grey2, text width=14.8cm, below = 1.cm of z] (II) {\myboxcontent{Numerical gIOC solver}{Statistical test positive on validation data set?}};
\draw[arr] ([yshift=-2pt]Ileft.south) to ([yshift=2pt,xshift=-2cm]II.north);
\draw[arr] ([yshift=-2pt]Iright.south) to ([yshift=2pt,xshift=2cm]II.north);

\node[mybox, fill = grey3, text width=14.8cm, below = 2.6cm of z] (III) {\myboxcontent{Positive result: identified optimality principle}{Objective $\phi$, constraints $\mathcal{Y}$, dynamics $f$, parameters $p^*$, states $x^*$, controls $u^*$}};
\draw[arr] ([yshift=-2pt]II.south) to ([yshift=2pt]III.north);

\node[mybox, fill = grey4, text width=14.8cm, below = 4.2cm of z] (IV) {\myboxcontent{Further use of optimality principle}{Biological insight; new testable hypothesis; forward optimal control; optimal inverventions}};
\draw[arr] ([yshift=-2pt]III.south) to ([yshift=2pt]IV.north);


\end{tikzpicture}
}
\caption{Proposed workflow to identify and use optimality principles via gIOC} \label{fig_gIOC_overview}
\end{center}
\end{figure*}
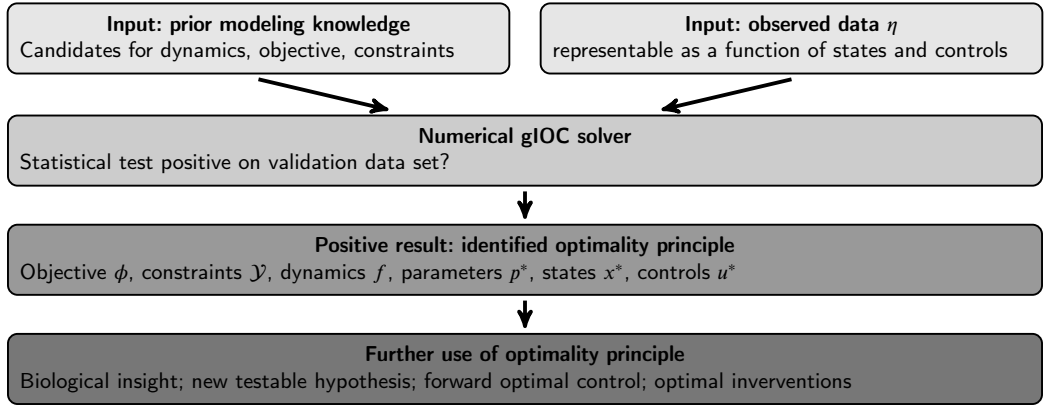

Note also that modeling choices and underlying assumptions have a strong \textbf{impact on the resulting optimal solutions} that are to be compared to observations. For example, for dynamics with linearly affine controls, $\dot{x}(t) = A(x(t),p) + B(x(t),p) u(t)$, the optimal structure is typically bang-bang for objective functionals of the type $\int_0^T \| u(\tau) \|_1 \; \mathrm{d}\tau$, i.e., all optimal controls take values at the lower or upper bound on $u(t)$, while there are singular arcs, i.e., values in the strict interior of the lower and upper bounds, for quadratic objective functionals such as $\int_0^T \| u(\tau) \|_2^2 \; \mathrm{d}\tau$. Similarly, the way dynamics and constraints are modeled influences the structure and quantitative values of independent and dependent variables.
Consider the car driving problem mentioned in the introduction and illustrated in Figure~\ref{fig_carSimple}. A simplified optimal control formulation would be: starting from rest, what is the optimal acceleration profile $a(t)$, between limits $a_M$ and $-a_M$, to go in a straight line from A to B in minimum time, stopping at B? The solution is bang-bang: maximum acceleration $a_M$ until the midpoint, then maximum decceleration $-a_M$ until the final time. This bang-bang policy is obviously not realistic because it involves abrupt switching between extreme values, which is often not feasible in real-world applications. But it is important to note that such an unrealistic bang-bang policy arises from our simplified modeling. If we consider a more refined car model, compare Figure~\ref{fig_carAdvanced}, where the dynamics, path constraints, and physical limitations of actuators and the engine are considered, one obtains different optimal control policies. The resulting velocities match experimental observations quite well \cite{Kehrle2011,dal2018minimum}, see also Figure~\ref{fig_carComparison}. A similar situation is found in the linear metabolic pathway considered by Klipp et al. \cite{Klipp2002-sg}. When simplified models are used, the solutions are bang-bang. However, when more realistic models are used, e.g., introducing kinetics for gene expression, the optimal policies may be smoother and capture the enzyme dynamics linked to the metabolic activity much better \cite{Oyarzun2009-ef,Bartl2010-vt,de2014global,Waldherr2015-rk}.

\subsubsection{A practical gIOC instance} \label{sec_special}

We summarize some of the assumptions we made along the way that are especially relevant in molecular systems biology to formulate a special case of \eqref{eq_generalIOC} that is closer to practical application.
We focus on dynamic systems and the particular case of ordinary differential equations \eqref{eq_ODEp}.
We assume that all (candidate) functions $\phi_i$, $f$, and $g_i$ are known, or that alternatively the symbolic regression task can be formulated using model parameters $p$.
We then define the following bi-level problem with differential states $x \in \mathcal{X}$, controls $u \in \mathcal{U}$, model parameters $p \in \mathbb{R}^{n_p}$, and convex multipliers $w = (w^\phi, w^g) \in \mathcal{W}$ 
\begin{equation} \label{eq_gIOC}
\hspace*{-0.2cm}
\begin{array}{rwc{21pt}lrcl}
\multicolumn{6}{l}{\displaystyle \min_{(p, w, x^*, u^*) \in \Omega_1} \; \| h(x^*, u^*) - \eta \| + R(p,w)} \\[2.7ex]
\multicolumn{6}{l}{\text{subject to}} \\
 & (x^*,u^*) \in & \multicolumn{4}{l}{ ~ \hspace*{0.01cm} \displaystyle \arg \hspace*{-0.5cm} \min_{(x, u) \in \Omega_2(p,w)} \; \sum_{i=1}^{n_\phi} w^\phi_i \; \phi_i[x,u,p]} \\[1.2ex]
 && \multicolumn{4}{l}{\text{ subject to }} \\
 &&& \dot{x}(t) &=& f(x(t), u(t), p) \\
 &&& 0 &\le& \displaystyle w^g_i \; g_i(x(t), u(t), p) \; \forall \; i \in [n_g]
\end{array}
\end{equation}
for $t \in [0,T]$ as a specific, but still very general gIOC of type \eqref{eq_generalIOC}. Here $\mathcal{X}$ and $\mathcal{U}$ are properly defined function spaces. The variables $w$ indicate which objective functions and which constraints are relevant in the inner problem. With $n_w = n_\phi + n_g$ we define the feasible set 
$$\mathcal{W} = \{ (w^\phi, w^g) \in [0,1]^{n_w}: \textstyle \sum_{i=1}^{n_\phi} w^\phi_i = 1, w^g \in \{0,1\}^{n_g} \}.$$
On the outer level, the feasible set is $\Omega_1 \subseteq \mathbb{R}^{n_p} \times \mathcal{W} \times \mathcal{X} \times \mathcal{U}$, while on the inner level $\Omega_2$ contains bounds, boundary conditions, mixed path and control constraints, and more involved constraints such as dwell time constraints. We have observational data $\eta \in \mathbb{R}^{n_\eta}$, a measurement function $h: \mathcal{X} \times \mathcal{U} \mapsto \mathbb{R}^{n_\eta}$, a regularization function with a priori knowledge on parameters and weights $R: \mathbb{R}^{n_p} \times \mathcal{W} \mapsto \mathbb{R}$ and candidate objective functionals $\phi_i: \mathcal{X} \times \mathcal {U} \times \mathbb{R}^{n_p} \mapsto \mathbb{R}$ and constraint functions $g_i: \mathcal{X} \times \mathcal {U} \times \mathbb{R}^{n_p} \mapsto \mathbb{R}^{n_g}$.

The unknown parts of the inner level optimal control problem are modeled using the convex multipliers $w$. As discussed above, they may also depend on time in the case of switched optimality principles and are only one particular way to write down the goal to identify objective function candidates contributing to Pareto optimality and to identify activity of constraints. 

\sreview
Problem formulation \eqref{eq_gIOC} is given in \textbf{continuous time}. It is also possible to formulate a similar discrete time version. Mathematically, for the choice of constant controls $u_k$ and
\begin{equation} \label{eq_MDP_CT}
F(x_k,u_k,p) := x_k + \int_{t_k}^{t_{k+1}} f(x(\tau), u_k \; \mathrm{d} \tau
\end{equation}
the solutions of the continuous time system
$$\dot{x}(t) = f(x(t), u_k, p), \quad x(t_k) = x_k, \quad t \in [{t_k, t_{k+1}}]$$
and of the discrete time system
\begin{equation} \label{eq_discretetime}
x_{k+1} = F(x_k, u_k, p)
\end{equation}
are identical in the sense of $x(t_k) = x_k$. Thus, variants of problem \eqref{eq_gIOC} that are based on \eqref{eq_discretetime} can be found in the literature.
Yet, there may be practical differences and community-dependent preferences. 
The discrete time version may be more convenient because the time grid $t_k$ can be chosen matching the data $\eta$ and many inverse reinforcement learning algorithms have been formulated for discrete time problems. 
Discrete-time formulations can be effectively used with mechanistic models as long as the discretization preserves the essential physical meaning and time scales of the underlying continuous processes through appropriate sampling rates and numerical methods. While the implementation may be discrete, the model's structure and parameters should still be derived from and interpreted in terms of the continuous-time physical processes they represent.
Solution properties such as bang-bang behavior are better discussed in continuous time.
The continuous time version is also closer to a direct mechanistic modeling and thus preferred in systems biology.  We will therefore continue with the notation of \eqref{eq_gIOC}.
\ereview

On the one hand, \eqref{eq_gIOC} is restrictive in the interest of a clearer presentation and might be further generalized, e.g., to multi-stage formulations involving partial differential-algebraic, delay differential, or stochastic differential equations. On the other hand, problem \eqref{eq_gIOC} is quite generic and allows, e.g., the consideration of switched systems, periodic processes, different underlying function spaces $\mathcal{X}$ and $\mathcal{U}$, and the usage of universal approximators such as neural networks as candidate functions. 
The proposed workflow is illustrated in Figure~\ref{fig_gIOC_overview}.

\subsection{Examples of forward control problems} \label{sec_examples}

\sreview
Molecular and cellular biological systems present unique challenges and characteristics when viewed through the lens of optimal control. Unlike engineered systems, biological processes are the result of evolution rather than deliberate design. In addition, these systems often operate under multiple competing objectives. For instance, cells must balance energy efficiency against speed of response, or growth rate against robustness to environmental changes. In this context, it is worth providing an overall description of such characteristics and challenges for the three key elements of an optimal control formulation discussed in Sections~\ref{sec_dynamics}--\ref{sec_constraints}: dynamics, objective function and constraints.

The \textbf{dynamics} in molecular systems biology typically involve mass-action kinetics, Michaelis-Menten enzyme kinetics, and Hill functions for describing regulatory interactions \cite{distefano2015dynamic,KlippTextBook2016}. These give rise to nonlinear ordinary differential equations, often with multiple time scales. A distinctive feature is that reaction rates are constrained to be non-negative, and species concentrations must remain within physiologically feasible ranges. Furthermore, many biological processes exhibit switch-like behavior, leading to dynamics that can be approximated by hybrid systems combining continuous and discrete elements.

Candidate \textbf{objective functions} in biological systems often reflect fundamental physical and evolutionary constraints. These may include minimization of energy expenditure (often expressed as ATP consumption), maximization of growth rate or biomass production, minimization of transition time between metabolic states, or maximization of robustness to perturbations \cite{Heinrich1991-gv,Heinrich1996-ni,Schuetz2007}. In many cases, these objectives must be considered simultaneously, leading to multi-objective optimization problems \cite{de2014global}. The relative weights between different objectives may themselves be state-dependent, reflecting the cell's ability to adapt its priorities based on environmental conditions.

Biological \textbf{constraints} are particularly complex and multifaceted \cite{Garland2022-tw}. At the molecular level, there are thermodynamic constraints on reaction directions and rates \cite{Bekiaris2023}, as well as physical constraints on enzyme concentrations and metabolite levels \cite{Klipp2002-sg,Bruggeman2020-fu,scott2023shaping}. Cellular machinery has limited capacity, leading to resource allocation constraints \cite{Erickson2017,Yabo2022}. For instance, ribosomes must be shared between different protein production processes. There are also temporal constraints, as many biological processes must complete within specific time windows, such as cell cycle phases. Additionally, homeostatic constraints require certain variables to be maintained within narrow ranges despite environmental fluctuations.
\ereview

\sreview
We focus on illustrating some fundamental concepts. For this, we start with a classical, non-biological example that is ideally suited to visualize the impact of objective function and constraints for the inner optimization problems in \eqref{eq_gIOC} on the optimal solution $y^*$ in two dimensions. As a next step, we consider a basic example from cell biology to highlight additional general issues arising in gIOC.
\ereview

\begin{figure*}[ht!!!]
\centering
  \begin{subfigure}{0.49\linewidth}
  \includegraphics[width=\linewidth]{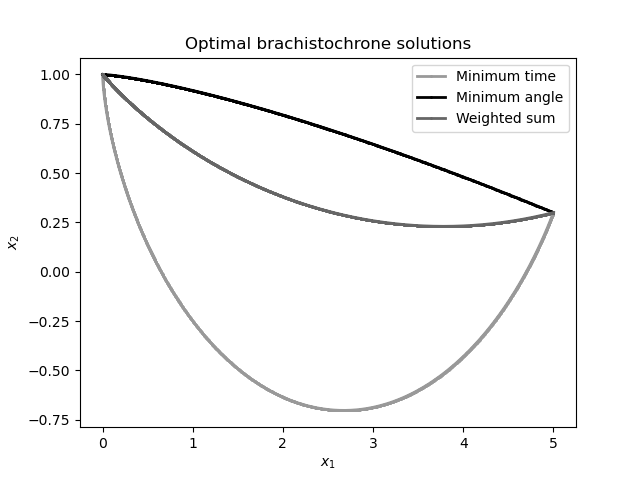}
  \caption{Optimal solutions for problems \eqref{eq_bractime}, \eqref{eq_bracangle}, and \eqref{eq_bracweighted} }\label{fig:bracA}
  \end{subfigure}
  \begin{subfigure}{0.49\linewidth}
  \includegraphics[width=\linewidth]{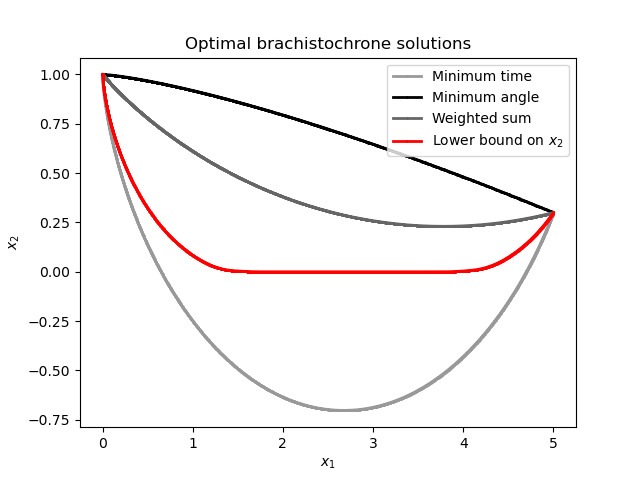}
  \caption{As \eqref{fig:bracA}, plus solution of constrained \eqref{eq_bracconstraint}} \label{fig:bracB}
  \end{subfigure}

  \begin{subfigure}{0.49\linewidth}
  \includegraphics[width=\linewidth]{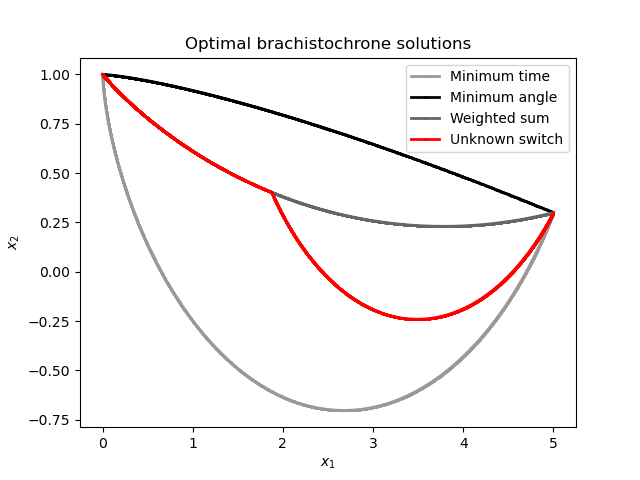}
  \caption{As \eqref{fig:bracA}, plus solution of switched \eqref{eq_bracswitchB}} \label{fig:bracC}
  \end{subfigure}
  \begin{subfigure}{0.49\linewidth}
  \includegraphics[width=\linewidth]{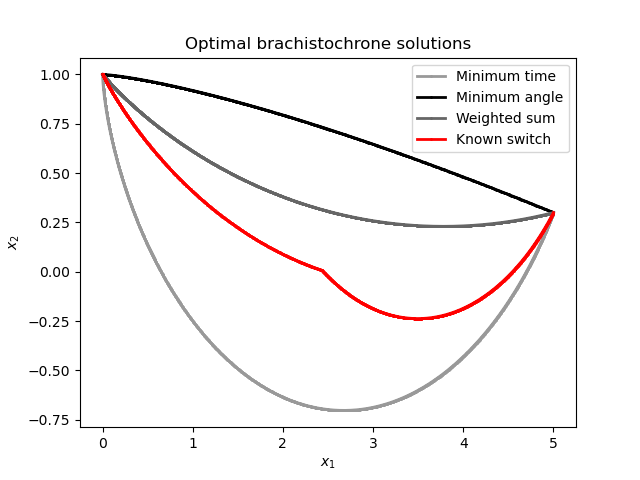}
\caption{As \eqref{fig:bracA}, plus solution of switched \eqref{eq_bracswitchA}} \label{fig:bracD}
  \end{subfigure}
\caption{Optimal solutions $y^*$ to brachistochrone problems (\ref{eq_bractime}--\ref{eq_bracswitchB}), projected onto $(x_1^*, x_2^*)$, indicating the difference in observable behavior (state trajectories) dependent on objective function, constraints, and scenario switching.\newline
\eqref{fig:bracA}: While the optimal solution to Bernoulli's minimum time problem \eqref{eq_bractime} is well known, other objective functions as in \eqref{eq_bracangle} result in different observable behavior. Pareto-optimal solutions with weighted objectives result in a compromise.\newline
\eqref{fig:bracB}: The additional solution to the $x_2(t) \ge 0$ constrained minimum time problem~\eqref{eq_bracconstraint} shows that an active constraint impacts the observed behavior. Note that the activity of the constraint for values of $1 \le x_1(t) \le 4$ also impacts the behavior for values $x_1(t) \le 1$ and $x_1(t) \ge 4$.\newline
\eqref{fig:bracC}: The solution for problem \eqref{eq_bracswitchB} modeling an external (unknown) switch at time $t=1$ and $x_1(1) \approx 2$. The switch from the weighted to a minimum time objective is clearly visible in the kink of the trajectory.\newline
\eqref{fig:bracD}: The solution for problem \eqref{eq_bracswitchA} modeling an internal (known) switch at time $t=1$ and $x_1(1) \approx 2.4$. Again, the switch from the weighted to a minimum time objective is clearly visible in the kink of the trajectory, but the solution differs from the solution to problem \eqref{eq_bracweighted} also for times $t \le 1$.
}
\label{fig_bracExamples}
\end{figure*}

\subsubsection{Brachistochrone} \label{sec_brachistochrone}

To acknowledge the contribution of variational calculus, often referred to as the ``parent of optimal control'' \cite{Kalman1964}, and for its intuitiveness, we examine the (non-biological) brachistochrone problem.
Formulated by Johann Bernoulli in 1696, it poses the question: ``Given two points A and B in a vertical plane, what is the curve traced out by a point acted on only by gravity, which starts at A and reaches B in the shortest time.''
The public challenge of Bernoulli in \textit{Acta Eruditorum} resulted in different solutions by Bernoulli himself (based on Fermat's principle), Jacob Bernoulli, Leibniz, de l’ Hopital, and Newton. In the 20th century, the consideration of practical problems and constrained solutions in different spaces resulted in the research areas (forward) optimal control and optimization. The historical perspective is well described in \cite{McDonough2022-vz,Nahin2021-lt,Sussmann1997-fu,Pesch2009-tq}.

We use a modern formulation of Bernoulli's problem that can be easily solved numerically, e.g., by first-discretize-then-optimize methods \cite{Biegler2010-ua}. We define the brachistochrone problem for the three differential states $x_1, x_2$, specifying the location of the particle in two dimensions, and the directional velocity $x_3$, using in \eqref{eq_gIOC} the right hand side function
\begin{align}
f_1(x, u) &= x_3 \cos(u) \\
f_2(x, u) &= - x_3 \sin(u) \\
f_3(x, u) &= G \sin(u)
\end{align}
on the time horizon $[0,T]$ with the unknown time $T$. To move the particle from a start position to an end position, we specify boundary constraints
\begin{align}
\Omega_2 := \left\{ x(0) = (0, 1, 0), x_1(T) = 5, x_2(T) = 0.3 \right\}.
\end{align}

The control $u \in \mathcal{U} := \{u: [0,T] \mapsto [-\pi/2, \pi/2]\}$ represents the angle of the plate the particle moves along.
With these definitions we can now state a version of the traditional minimum time brachistochrone problem,
\begin{align} \label{eq_bractime}
\min_{T, u \in \mathcal{U}, x \in \mathcal{X}} \; T
\end{align}
as well as alternative \textbf{objectives} minimizing the integrated square angle,
\begin{align} \label{eq_bracangle}
\min_{T, u \in \mathcal{U}, x \in \mathcal{X}} \; \int_0^T u(\tau)^2 \; \mathrm{d} \tau
\end{align}
a weighted combination of both
\begin{align} \label{eq_bracweighted}
\min_{T, u \in \mathcal{U}, x \in \mathcal{X}} \; w_1 T + w_2 \int_0^T u(\tau)^2 \; \mathrm{d} \tau,
\end{align}
e.g., with weights $w_1 = \frac{1}{4}$ and $w_2 = \frac{3}{4}$,
and a \textbf{state-constrained} problem
\begin{align} \label{eq_bracconstraint}
\min_{T, u \in \mathcal{U}, x \in \mathcal{X}} \; T \text{ subject to } x_2(t) \ge 0 \; \forall \; t \in [0,1].
\end{align}
We also consider two scenarios with a \textbf{switch} in the objective function at time $t_1=1$ from objective~\eqref{eq_bracweighted} to \eqref{eq_bractime}, i.e., with $w_1 = \frac{1}{4}$ and $w_2 = \frac{3}{4}$ for $t \in [0,t_1]$ and with $w_1=1$, $w_2=0$ for $t \in [t_1,T]$. 
The first problem is formulated such that the optimization starts at time $1$ with a partial solution $\hat{x}(1)$ of the previous optimization problem \eqref{eq_bracweighted}
\begin{align} \label{eq_bracswitchB}
\min_{T, u \in \mathcal{U}, x \in \mathcal{X}} \; T \text{ subject to } x(1) = \hat{x}(1)
\end{align}
and corresponds to an environmental change occuring at time $t_1=1$.
The second one,
\begin{align} \label{eq_bracswitchA}
\min_{T, u \in \mathcal{U}, x \in \mathcal{X}} \; \frac{1}{4} + \frac{3}{4} \int_0^1 u(\tau)^2 \mathrm{d} \tau + T,
\end{align}
takes the switch of the objective function at time $1$ into account when optimizing. 
The optimal solutions for all six problems projected to the $(x_1,x_2)$ space are plotted in Figure~\ref{fig_bracExamples}.
Associated optimal values $T^*$ of (\ref{eq_bractime}--\ref{eq_bracswitchA}) are
1.44,
$\infty$ (if unbounded),
1.83,
1.51, 
2.59, and
8.67,
respectively.

The main task of gIOC is the inverse problem: given the solutions shown in Figure~\ref{fig_bracExamples}, and assuming corresponding observations with measurement noise, is it possible to automatically and systematically derive the optimal control problem that was optimized?
For the brachistochrone problem a practical analogy is unrealistic, one should rather think of an intellectual puzzle.
The main take-away of this brief study is an \textbf{intuition} about how different objectives, constraints, and assumptions on switching the optimality principle during the observed time horizon may impact the outcome.

\subsubsection{A metabolic pathway problem} \label{sec_metabolic}

To acknowledge the standard metabolic pathway problem in systems biology \cite{Klipp2002-sg}, already illustrated in Figure~\ref{fig:panelLinearPathway}, we use its formulation as an optimal control \cite{Oyarzun2009-ef,Bartl2010-vt} to \textbf{illustrate further challenges} in gIOC. 
We consider a 3-step linear metabolic pathway with mass action kinetics. 
The differential states $x$ represent metabolite concentrations, the time-dependent control functions $u$ represent enzyme concentrations, and all kinetic parameters in the mass action expressions are assumed to be $1$ for simplicity. The differential equations are formulated over the time horizon $[0,T]$ as
\begin{align} \label{eq_metaboliclinear}
f(x, u) & = (u_1 - x_1 u_2, x_1 u_2 - x_2 u_3, x_2 u_3)
\end{align}
with fixed initial and terminal values of the metabolite concentrations as boundary conditions
\begin{align}
\Omega_2 := \left\{ x(0) = (0, 0, 0), x_3(T) = 0.9 \right\}.
\end{align}
In the literature, also \textbf{different dynamics} have been investigated \cite{de2014global}, such as Michaelis-Menten kinetics
\begin{align} \label{eq_metabolicMM1}
f_1(x, u) & = u_1 - \frac{2 x_1}{1 + x_1} u_2 \\
f_2(x, u) & = \frac{2 x_1}{1 + x_1} u_2 - \frac{2 x_2}{1 + x_2} u_3 \\
f_3(x, u) & = \frac{2 x_3}{1 + x_3} u_3). \label{eq_metabolicMM3}
\end{align}

\begin{figure*}
\centering
  \begin{subfigure}{\linewidth}
  \includegraphics[width=0.44\linewidth]{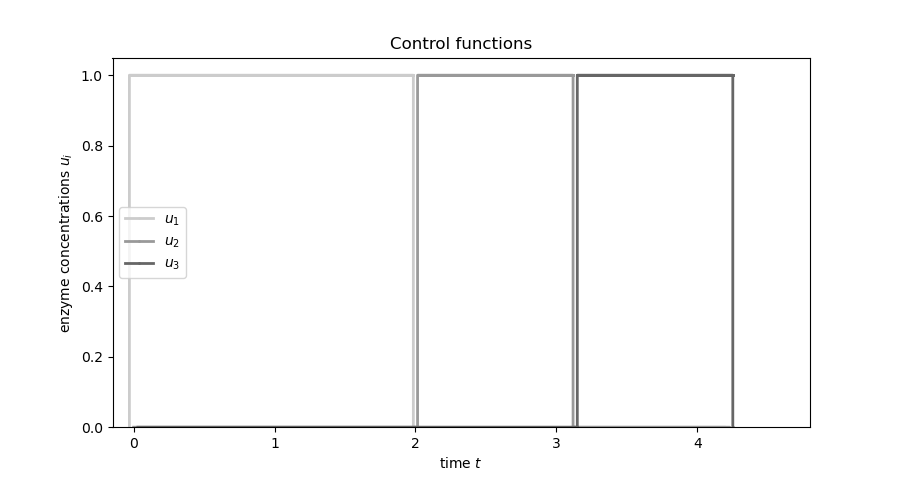} \hfill
  \includegraphics[width=0.44\linewidth]{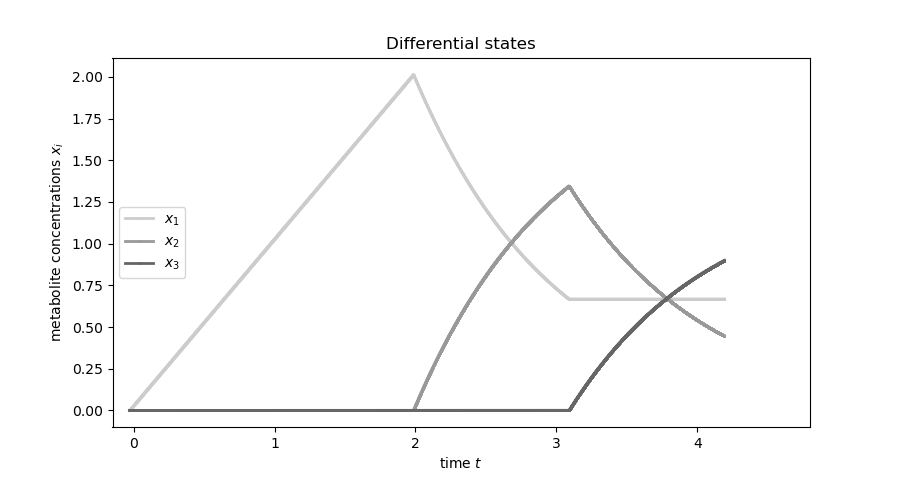}
  \caption{Optimal solution for problems \eqref{eq_metabolictime} and \eqref{eq_metabolicu}}\label{fig:metabolicA}
  \end{subfigure}
%
  \begin{subfigure}{\linewidth}
  \includegraphics[width=0.44\linewidth]{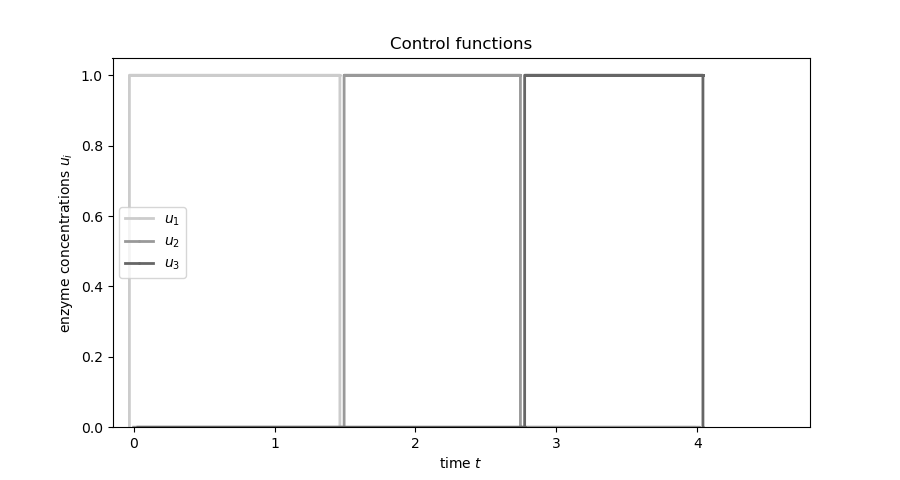} \hfill
  \includegraphics[width=0.44\linewidth]{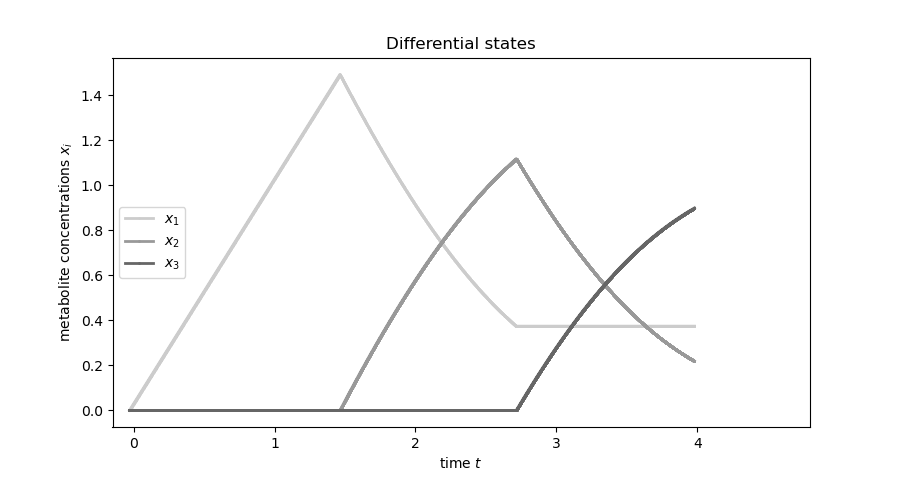}
  \caption{Optimal solution for problems \eqref{eq_metabolictime}, but with Michaelis-Menten kinetics (\ref{eq_metabolicMM1}-\ref{eq_metabolicMM3}).}\label{fig:metabolicD}
  \end{subfigure}
%
  \begin{subfigure}{\linewidth}
  \includegraphics[width=0.44\linewidth]{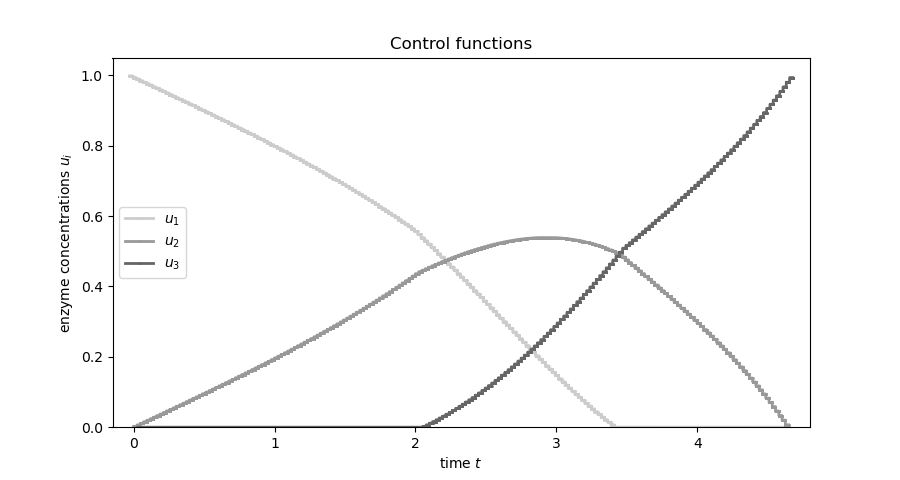} \hfill
  \includegraphics[width=0.44\linewidth]{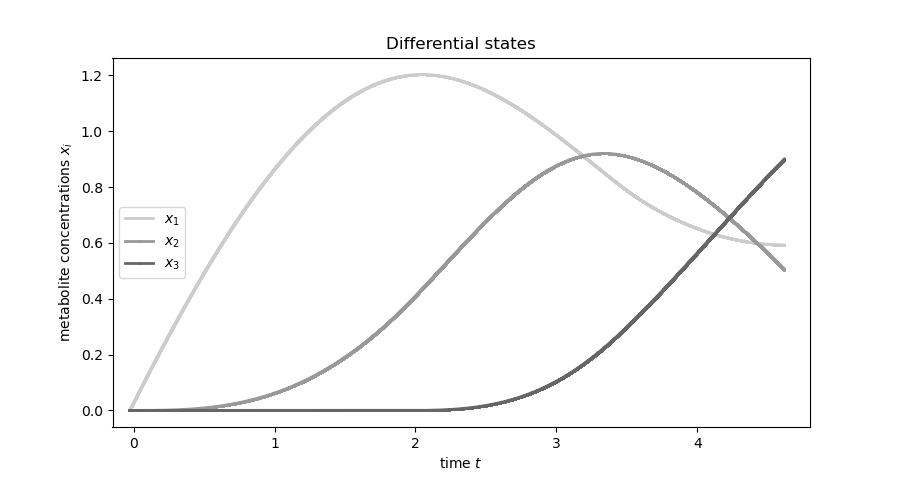}
  \caption{Optimal solution for problem \eqref{eq_metabolicusquare} }\label{fig:metabolicB}
  \end{subfigure}
  \begin{subfigure}{\linewidth}
  \includegraphics[width=0.44\linewidth]{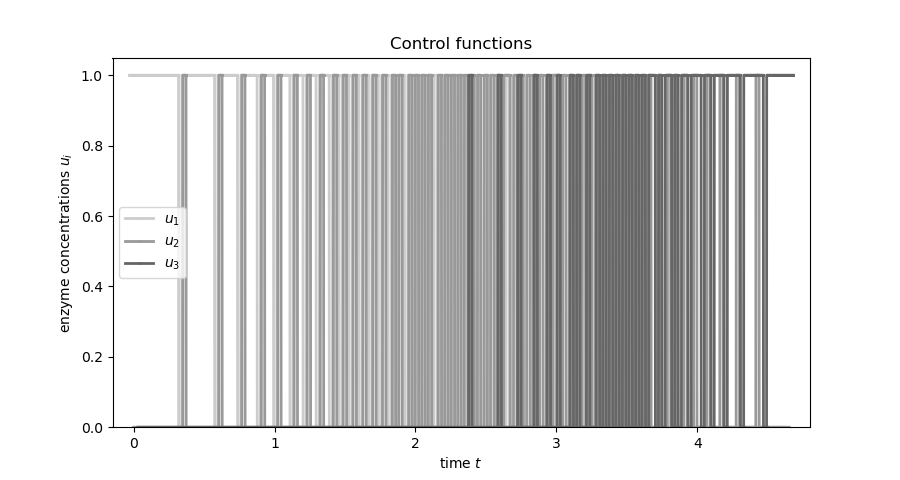} \hfill
  \includegraphics[width=0.44\linewidth]{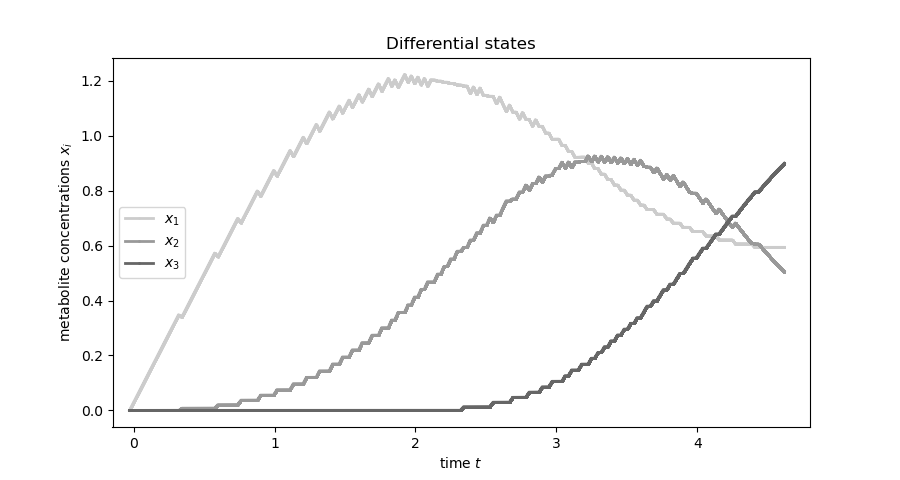}
  \caption{Solution with different controls, but similar states compared to (\ref{fig:metabolicB}) }\label{fig:metabolicC}
  \end{subfigure}
\caption{Visualization of control functions and corresponding differential states of the metabolic pathway problem.\newline
(\ref{fig:metabolicA}): solution of the time minimal objective function~(\ref{eq_metabolictime}) as already illustrated in Figure~\ref{fig:panelLinearPathway}. Due to the particular bang-bang structure it is also optimal for the enzyme activiation minimizing function \eqref{eq_metabolicu}.
(\ref{fig:metabolicD}): modifying the assumed ODE model from \eqref{eq_metaboliclinear} to (\ref{eq_metabolicMM1}-\ref{eq_metabolicMM3}) results in a structurally similar optimal control, but different differential states $x^*$ (note the different scale on the y-axis).
(\ref{fig:metabolicB}): Optimal solution for the weighted sum problem \eqref{eq_metabolicusquare} with structurally different optimal controls. The reduction in $\int_0^T u^2_1(\tau) + u^2_2(\tau) + u^2_3(\tau) \; \mathrm{d}\tau$ leads also to a reduction in use of intermediate metabolites $x_1$ and $x_2$. It is balanced by an increased time $T^*$ to reach $x_3(T^*)=0.9$.
(\ref{fig:metabolicC}): The differential states from (\ref{fig:metabolicB}) can be arbitrarily closely approximated, although the chattering bang-bang control functions are structurally different from those in (\ref{fig:metabolicB}). This shows the importance of choosing the observables for gIOC, especially considering that the variables $y^*$ can typically only be observed with noise.
}
\label{fig_metabolicExamples}
\end{figure*}

The control functions 
$$u \in \mathcal{U} := \{u: [0,T] \mapsto [0, 1]^3, \sum_{i=1}^3 u_i(t) \le 1\}$$ 
need to consider an inequality constraint. Limitations due to molecular crowding impose an upper bound on the maximum total concentration of enzymes (controls) at any given time. This constraint has a significant impact on $y^*$, similar to the study of problem \eqref{eq_bracconstraint} above. 

Optimal control of this and related systems has been studied extensively in the literature, e.g., \cite{Oyarzun2009-ef,Bartl2010-vt,de2014global,nimmegeers2016dynamic,Tsiantis2018}. We use it here to illustrate some \textbf{identifiability challenges} in gIOC. For this, we consider different candidate objective functionals $\phi_i$, namely the final time to reach $x_3(T)=0.9$, i.e.,
\begin{align} \label{eq_metabolictime}
\min_{T, u \in \mathcal{U}, x \in \mathcal{X}} \; T
\end{align}
the overall amount of enzymes
\begin{align} \label{eq_metabolicu}
\min_{T, u \in \mathcal{U}, x \in \mathcal{X}} \; \int_0^T u_1(\tau) + u_2(\tau) + u_3(\tau) \; \mathrm{d}\tau
\end{align}
and a weighted sum involving a squared penalization of enzyme activity,
\begin{align} \label{eq_metabolicusquare}
\min_{T, u \in \mathcal{U}, x \in \mathcal{X}} \; \frac{1}{2} T + \frac{1}{2} \int_0^T u^2_1(\tau) + u^2_2(\tau) + u^2_3(\tau) \; \mathrm{d}\tau
\end{align}
that we mainly investigate for illustration as it leads mathematically to a singular solution with optimal values $u^*_i(t) \in (0,1)$.

Figure~\ref{fig:metabolicA} shows trajectories $x$ and $u$ illustrating that different objective functions $\phi_i$ may result in the same observables $y^*$. We emphasize in Figure~\eqref{fig:metabolicD} that different assumptions may result in structurally similar optimal regulation but different differential states, and in  Figure~\eqref{fig:metabolicC} that apparently different regulations $u$ can result in arbitrarily similar state behavior $x$.

\subsection{Methods for gIOC} \label{sec_methods}

\sreview
Solving \eqref{eq_generalIOC} in full generality is a very difficult task. 
\ereview
However, algorithms for special cases have been investigated in the literature, which can serve as building blocks for a future gIOC methodology. 
\sreview
We discuss the relations to system identification and interpretation in Sections~\ref{sec_modelidentification} and \ref{sec_modelinterpretation} and to inverse reinforcement learning in Section~\ref{sec_irl}, before we address the bi-level structure of \eqref{eq_generalIOC} in \ref{sec_bilevel}, how to solve forward control problems in \ref{sec_innerproblem}, and recent extensions involving inference of constraints or multi-phase objectives in \ref{sec_extensions}. We discuss identifiability issues in Section~\ref{sec_identifiability} and illustrate the main concepts with examples in \ref{sec_gIOCexamples}.
\ereview

\FloatBarrier

\subsubsection{Model identification} \label{sec_modelidentification}
\sreview 
A gIOC subproblem that is very relevant and actively researched is model identification (or system identification), the process of building mathematical models of dynamic systems using measured data \cite{Ljung2010,distefano2015dynamic}. Dynamic systems theory is a powerful tool for modeling and understanding complex systems in nature and engineering. Dynamic systems theory provides the language, concepts, and framework for describing how systems evolve over time, while model identification is the practical process of finding a specific mathematical model within that framework that matches a real-world system.  
Model identification involves determining both the structure (the form of the equations) and parameters of a model that best explain observed input-output behavior of the system. This process typically includes selecting model candidates, designing informative experiments, collecting data, estimating parameters, and validating the resulting model against independent data \cite{walter1997identification}. Dynamic model identification distinguishes between mechanistic "white-box" models derived from first principles for physical insight, and empirical "black-box" models that are purely data-driven for flexibility at the cost of interpretability. In practice, "grey-box" models offer a pragmatic compromise, integrating known theory with data-driven components to balance predictive accuracy with scientific meaning.

By encoding the underlying physical and (bio)chemical principles, \textbf{mechanistic models} provide the crucial advantages of offering deep scientific insight and the power to make reliable predictions far beyond the original experimental conditions. However, developing mechanistic dynamic models of biological systems is particularly challenging, as we cannot rely on first principles in the same way as in physics. Consequently, systematic model development is one of the key open problems in mathematical biology. 

Model discovery, i.e., the symbolic reconstruction of equations from data, is a highly desirable approach to automating the development of dynamic models. A number of different statistical and machine learning frameworks have been considered for model discovery, including symbolic regression, grammar-based methods, sparse regression, neural networks, Gaussian process regression, and Bayesian approaches, as reviewed in \cite{Ghadami2022-fj,Brunton2019-wp,Dzeroski2008-ja,camps2023discovering,metzcar2024review}. The sparse identification of nonlinear dynamics (SINDy) algorithm has been particularly successful \cite{Brunton2016-vw,Mangan2016-ge}. Other approaches solve the symbolic regression problem via mixed-integer optimization \cite{Cozad2018-lb,Kim2023-bt,Bertsimas2023-ge,Austel2017-xt}.

There are many open research questions in model discovery, including its generalization for the concurrent identification of objective function, constraints, and dynamics in inverse problems. For such ambitious goals, the availability and quality of data are essential. Here, \textbf{identifiability and observability} are crucial structural properties \cite{Villaverde2014,wieland2021structural}, playing also a key role in the general problem of data-driven model discovery \cite{massonis2023distilling}. Identifiability and observability are two fundamental properties of dynamic models, each with a theoretical and a real-world instance. Identifiability concerns whether model parameters can be uniquely determined from data; structural identifiability is the theoretical possibility of doing so with perfect data based on the model's equations, while practical identifiability assesses if parameters can be reliably estimated from actual, noisy experimental data \cite{wieland2021structural}. Similarly, observability addresses whether a system's internal states can be deduced from its outputs; structural observability is the theoretical possibility of this reconstruction, whereas practical observability is the ability to effectively estimate those states in the face of real-world measurement limitations \cite{villaverde2019observability}.

Consequently, a model might be structurally identifiable but practically unidentifiable due to measurement noise, insufficient data quality or quantity, high parameter correlation, or poor experimental design. This distinction is crucial in biological systems where measurements are often sparse, noisy, and expensive to obtain. Furthermore, practical identifiability analysis can guide experimental design by revealing which measurements and perturbations are most informative for parameter estimation, and can help determine confidence intervals for parameter estimates. The interplay between these two forms of identifiability is particularly important in complex biological models where parameters often have mechanistic interpretations, and their accurate estimation is crucial for model prediction and control \cite{Villaverde2014,distefano2015dynamic}.

When considering the more general problem of inverse optimal control (IOC), these identifiability issues take on new dimensions. While methods have been proposed for testing structural identifiability in standard models, even with time-varying inputs \cite{Villaverde2019-xp}, to the best of our knowledge there is no procedure for testing the equivalent of this property for IOC problems. Such issues have, however, been detected by inspection \cite{Tsiantis2018}, suggesting they will also be present in generalized IOC frameworks. Moreover, even with structural identifiability ensured, practical identifiability remains a key hurdle. In the context of IOC, non-informative data will result in the non-unique identification of the entire optimality principle—including the cost function, constraints, and controls—not just the dynamic model parameters.
\ereview

\sreview
\subsubsection{Model interpretability} \label{sec_modelinterpretation}
Another key property is model interpretability \cite{massonis2023distilling,Choudhury2025,Brunton2025}, i.e. the ability to understand and explain a model's structure and behavior in terms of known biological mechanisms and physical principles. An interpretable model uses mathematical expressions that correspond to real biological processes (such as enzyme kinetics, binding interactions, or regulatory relationships) rather than abstract mathematical expressions that may fit the data but lack biological meaning. This means that not only should the model's predictions be accurate, but 
its structure and parameters should provide insights into the underlying biological mechanisms, allowing scientists to connect mathematical terms with specific molecular or cellular processes. 

The interplay between interpretability, identifiability, and observability in biological modeling presents a complex challenge that lies at the heart of mechanistic model development \cite{Massonis2021}. A mechanistically interpretable model should reflect actual biological processes. However, the pursuit of 
mechanistic meaning must be balanced with the mathematical properties of identifiability and observability to ensure the model's practical utility. While identifiability and observability are essential properties for the model to 
have predictive power and provide meaningful biological insights, there can be tension between these properties and mechanistic interpretability. For instance, a mechanistically meaningful model might include detailed enzyme kinetics with multiple parameters, but this complexity could lead to practical unidentifiability. Conversely, a mathematically simpler model that ensures identifiability might 
lack biological meaning or fail to capture important mechanistic details.

The challenge is to find representations that maintain mechanistic interpretability while ensuring identifiability and observability. This might involve careful reparametrization of mechanistic terms, lumping of parameters, or reduction of model complexity while preserving essential biological features \cite{Massonis2021}. For example, a complex Michaelis-Menten mechanism might need to be simplified to a first-order reaction under certain conditions, or multiple sequential steps might need to be represented as a single effective process. The key is to make such simplifications in ways that preserve the biological meaning of the model while ensuring its mathematical properties allow 
for meaningful parameter estimation and state prediction. This balance is particularly important in the context of automated model discovery methods, where the challenge is to encode both mechanistic constraints and mathematical requirements into the model selection process. Initial progress toward addressing these challenges has been reported in \cite{massonis2023distilling} for the case of automatic model discovery, but extending these results to the general inverse optimal control framework remains an open research direction.
\ereview

\sreview
\subsubsection{Inverse Reinforcement Learning and IOC} \label{sec_irl}

Inverse Reinforcement Learning (IRL) is a machine learning technique that infers an agent's underlying reward function by observing its behavior or demonstrations. Instead of being given a reward, IRL works backward to discover what objective the agent was optimizing to produce its observed actions \cite{ng2000algorithms,arora2021survey}.
This general definition of inferring underlying objectives from observed behavior is very similar to the one of IOC, up to the point that both concepts are sometimes used interchangeably in the literature, e.g., ``Inverse optimal control, also known as inverse reinforcement learning, \dots '' \cite{jin2021inverse,levine2012continuous}.
\ereview

An extensive historical review of the two approaches, tracing back to early work of Kalman on closed loop IOC for linear-quadratic systems \cite{Kalman1964}, is provided in \cite{Ab_Azar2020-vt}. The authors also classify and discuss similarities and differences between IOC and IRL. However, they view IOC like Kalman, as \sreview mainly \ereview related to stabilizing feedback controls, which is not our focus, as discussed in Section~\ref{sec_further}. 

\sreview
In our view, and closer to another IRL survey \cite{Adams2022}, the \textbf{main differences} are linked to
terminology and historical context (IOC has roots in classical control theory, where systems are typically deterministic and continuous-time; IRL emerged more from the machine learning community),
underlying system models (IOC traditionally assumes a state-space model; IRL explicitly assumes a Markov Decision Process (MDP), accounting for stochastic environments and agent experience), and
computational approaches (IOC is based on optimization and control theory, IRL often applies stochastic optimization or machine learning algorithms).
An additional practical difference arises from the availability of data. 
Modeling with MDP or deep neural networks in IRL requires a large amount of data to avoid overfitting \cite{ng2000algorithms}. Many RL concepts, such as those in games like chess or Go, rely on the availability of producing data via simulation.
In contrast, in molecular biological applications, data is often sparse and possibly expensive to obtain. Also, interpretability of results may be more of an issue when compared to other fields of application, such as robotics or games.
Therefore mechanistic or hybrid models using differential equations may be better suited in this context.

Notwithstanding, we share the impression that ``the fields of IRL and IOC have essentially merged and become interchangeable''  \cite{Adams2022}, if general concepts and algorithmic ideas are concerned. This convergence is driven by model identification as outlined in Section~\ref{sec_modelidentification} and by a trend towards ``hybrid modeling'': mechanistic ordinary differential equation (ODE) models are often enhanced with data-driven surrogate submodels (UDE) on the one hand \cite{rackauckas2020universal}, and physics-informed universal approximators (PINN) \cite{raissi2019physics} have become an important field of research on the other hand. This convergence of prediction models is stimulating similar algorithmic developments for both IOC and IRL.

\vspace*{-0.2cm}
\begin{center}    
\begin{tikzpicture}
    \node at (-0.385,0.8) {\textbf{Converging models}};
    \node at (2.825,0.8) {\textbf{Algorithms}};
    \node[draw, rectangle, rounded corners, inner sep=5pt, fill=grey1, minimum width=2cm] (ode) at (0,0) {ODE};
    \node[draw, rectangle, rounded corners, inner sep=5pt, fill=grey1, minimum width=2cm] at (0,-0.8) {UDE};
    \node[draw, rectangle, rounded corners, inner sep=5pt, fill=grey1, minimum width=2cm] at (0,-1.6) {\ldots};
    \node[draw, rectangle, rounded corners, inner sep=5pt, fill=grey1, minimum width=2cm] at (0,-2.4) {PINN};
    \node[draw, rectangle, rounded corners, inner sep=5pt, fill=grey1, minimum width=2cm] at (0,-3.2) {MDP};
    \node[draw, rectangle, rounded corners, inner sep=5pt, fill=grey2, minimum width=2cm] at (3,0) {IOC};
    \node[draw, rectangle, rounded corners, inner sep=5pt, fill=grey2, minimum width=2cm] at (3,-1.6) {gIOC};
    \node[draw, rectangle, rounded corners, inner sep=5pt, fill=grey2, minimum width=2cm] at (3,-3.2) {IRL};
    \draw[->, thick, bend right] (-1.3,0) to (-1.3,-1.4) {};
    \draw[->, thick, bend left] (-1.3,-3.2) to (-1.3,-1.8) {};
    \draw[->, thick, bend left] (4.2,0) to (4.3,-1.4) {};
    \draw[->, thick, bend right] (4.2,-3.2) to (4.3,-1.8) {};
\end{tikzpicture}
\end{center}    
\vspace*{-0.1cm}

So, in theory and by making use of \eqref{eq_MDP_CT} to represent continuous time dynamics as a special case of a MDP, all IRL algorithms can also find application in IOC.
IRL methods, such as maximum entropy \cite{ziebart2008maximum} or GAIL, have been surveyed in \cite{Adams2022}.
It is an open and fascinating research question for which types of underlying prediction models these algorithms can be applied successfully in practice.

We believe that the \textbf{unique demands of systems biology} (such as the need for interpretable mechanistic differential equation models, the vastly different scale and quality of biological training data, and the availability of efficient nonlinear optimal control solvers) necessitate a specialized approach to gIOC algorithm development. This effort must extend beyond a straightforward adaptation of IRL methods.

\ereview

\subsubsection{Algorithmic approaches to the bi-level problem} \label{sec_bilevel}

Our interpretation of gIOC is inspired by and closely aligned with the work of \cite{Albrecht2011,Mombaur2017-ur,Mombaur2009-ur,Hatz2012-ta}. The authors formalized a bi-level formulation for the special case of fully known dynamics, switching structure, and active constraints.

Conceptually, \eqref{eq_generalIOC} and \eqref{eq_gIOC} represent bi-level optimization problems. On the \textbf{outer level} of \eqref{eq_gIOC}, a norm $\| \cdot \|$ and the regularization term $R$ define a data fit (regression) problem for observations $\eta$. These components relate to prior knowledge and statistical assumptions forming an objective function $$\displaystyle \min_{(p, w, x^*, u^*) \in \Omega_1} \; \| h(x^*, u^*) - \eta \| + R(p,w)$$ for model parameters $p$, convex multipliers $w$ that quantify the relevance of objective function candidates for the observed process, and states $x^*$ and controls $u^*$ that are optimal for this process. 
On the \textbf{inner level}, the gIOC problem is constrained by a potentially nonconvex optimal control problem (OCP). Solving this OCP yields $(x^*, u^*)$, which is necessary for evaluating the outer level objective function.

In \cite{Albrecht2011,Mombaur2017-ur,Mombaur2009-ur,Hatz2012-ta}, two possible methodological approaches for IOC have been discussed. The first one is to consider the inner optimal control problem as a black box, evaluated with a forward optimal control solver (from now on referred to as \textbf{two-level approach}), as described in e.g. \cite{Mombaur2009-ur}. On the outer level, a derivative-free method \cite{conn2009introduction} can be applied. Also, the a posteriori comparison of all solutions on the Pareto front, calculated using methods from multi-objective optimization, to the given data as suggested in \cite{Tsiantis2018} with case studies in biology, can be considered a two-level approach. The advantage here is the availability of the whole Pareto set for posteriori analysis of solution candidates. A disadvantage, however, is the probably higher computational cost to calculate the full Pareto front, especially if it depends on additional model parameters $p$.

%
The second approach involves reformulating the inner problem using conditions of optimality. This results in a one-level optimization problem (thus, from now on \textbf{one-level approach}). For example, as suggested in \cite{Albrecht2011} in the simpler case of equality constraints on the inner level, first order necessary conditions can be used. 
A comparison of these approaches on selected benchmark problems can be found in the PhD thesis of Hatz \cite{Hatz2012-ta}.
Simplifying and summarizing, one can expect the two-level approach to work robustly and, to a certain degree, with existing software. However, for larger problems the slow convergence rate will cause a problem. The one-level approach may offer advantages regarding computation time but is more involved from theoretical and practical point standpoints. A reformulation using first order necessary conditions of optimality does not guarantee to find a global optimum and is hindered by the non-smoothness of the complementarity conditions. An overview of approaches and computational aspects of bi-level optimization, involving integrality and uncertainty, can be found in the surveys of \cite{Kleinert2021-vg,Beck2023-de}. Particularly noteworthy are relaxations and reformulations as mathematical programs with equilibrium constraints \cite{Kim2020-hi} and concepts such as lifting \cite{Hatz2013-rr}. Another alternative for the special case of polynomial systems involves relaxing the inner control problem using a measure problem, which transforms it into a moment problem and transferring it to linear matrix inequalities, as discussed in \cite{Pauwels2016-gj}. 
\sreview
As mentioned above, there is also a great number of algorithms that have been suggested in the context of IRL that might be applicable to IOC as well.
Particularly interesting might be primal-dual algorithms that iterate between the outer and the inner optimization problems. We refer to the comprehensive IRL surveys \cite{ng2000algorithms,arora2021survey,Ab_Azar2020-vt,Adams2022} for further information.
\ereview

\subsubsection{Solving the inner optimization problem} \label{sec_innerproblem}

In both (one-level and two-level) approaches the efficient solution of (forward) optimal control problems is necessary. 
One distinguishes different classes of algorithms. Dynamic programming \cite{Bellman1954} and methods based on transformation into moment problems, compare the references in \cite{Sager2015}, both provide globally optimal solutions. However, they both signiticantly suffer from the \textit{curse of dimensionality} and are thus, despite ongoing efforts to reduce complexity, often computationally infeasible for problems with high-dimensional $y$.

Local solutions to optimal control problems can be calculated using indirect (\textbf{first-optimize then-discretize}) methods. Here the basic idea is to apply the necessary conditions of optimality, Pontryagin's maximum principle \cite{Pontryagin1962}, to the optimal control problem. The resulting boundary value problem is then solved with Newton-type methods. One advantage of this approach is that explicit formulae for feedback laws $u^*(x(t))$ and formal proofs of optimality might be derived analytically.
Alternatively, direct (\textbf{first-discretize then-optimize}) methods rely a finite-dimensional approximation of control functions and path constraints, resulting in a nonlinear programming problem (NLP) in finite dimension. 
Optimization algorithms are usually iterative. Starting from an initial guess $y^0$, they result in iterates $y^k$ converging towards a solution of a system of equality and inequality constraints specifying conditions of optimality. A well-known example are the necessary first order Karush-Kuhn-Tucker conditions. As an important advantage, direct methods have been shown to converge for more initial guesses $y^0$ when compared to first-optimize then-discretize methods applied to boundary value problems resulting from Pontryagin's maximum principle. Moreover, they can leverage the sophisticated implementation of established NLP solvers such as \texttt{ipopt} \cite{Waechter2006} and can be flexibly applied without requiring initial guesses concerning the optimal switching structure. 
Therefore, we consider direct collocation and direct multiple shooting as surveyed in \cite{Betts2001,Diehl2002b,Biegler2010-ua} as the most promising approaches for general optimal control problems arising on the inner level of \eqref{eq_gIOC}.

Often, controls are restricted to integer values, for instance, due to inhibition, which further increases the complexity. Control problems with such integrality constraints are referred to as \textbf{mixed-integer optimal control}, switched systems, or hybrid systems. Particularly promising for efficiently solving inner-level problems is the partial outer convexification approach suggested in \cite{Sager2012-vd}, as it allows the error-controled relaxation of the integrality constraints. This approach is particularly helpful when the observable $\eta$ comprises states. 
For applications of mixed-integer optimal control in cell signaling, refer to \cite{Lebiedz2005-mo,Slaby2007-tr}.

An overview and a comparison of different solvers, along with a discussion of local/global and single/multicriteria optimality in the context of systems biology, can be found in \cite{de2014global,Tsiantis2020-nf}.

\sreview
\subsubsection{IOC Extensions} \label{sec_extensions}

Different extensions are necessary to solve the gIOC problem \eqref{eq_gIOC} in full generality.
\ereview

As discussed in Section~\ref{sec_constraints} and exemplified in the two examples, it may be necessary to infer if a \textbf{constraint} $g_i(y^*) \ge 0$ is active and hence influencing the optimal solution $y^*$.
As a first remark, in optimal control one distinguishes between \textit{hard constraints} and \textit{soft constraints}, e.g., \cite{kerrigan2000soft}. Soft constraints enter the optimization as a penalty term in the objective. The penalty parameter determines the extent to which certain values of $y$ become less probable as optimal solutions. In contrast, hard constraints make values $y$ rigorously infeasible for the considered problem if $g_i(y) < 0$. 
In certain cases, hard constraints can be formulated as soft constraints, refer \cite{kerrigan2000soft} for reformulations and further references. This implies that methods used to infer objective functions (from IOC) could then also solve gIOC problems with unknown constraints. However, reformulating in this manner is not always straightforward, and usually the exact penalty parameter is unknown. Therefore we believe that numerical methods capable of inferring active hard constraints exactly must be developed.

The numerical methods will depend strongly on how constraints are modeled and which general approach to solve \eqref{eq_gIOC} is chosen. As mentioned above, the formulation $0 \le w_i g_i(y)$ with binary variables $w_i$ is only one possibility. Other options include continuous parameters or vanishing constraints, that might lead to specialized methods for complementarity constraints.

\sreview
First steps in the direction of a concurrent inference of objective function and constraints have been taken in the robotics and model predictive control community \cite{Menner2020}. Here, often a safe operation is paramount and dealing with a priori unknown obstacles and constraints is practically relevant \cite{Rickenbach2024}. An algorithm for learning parametric constraints from locally-optimal demonstrations \cite{chou2020learning} and the Safe Pontryagin Differentiable Programming paradigm \cite{jin2021safe} have been suggested. Both leverage necessary conditions of optimality, similar to the one-level approach mentioned in the previous section.
A different approach was investigated in \cite{ren2025inverse}. Here the learning of feasible regions is addressed for the special case of linear objectives, using block coordinate descent and mixed-integer linear programming.

The possible \textbf{switching of objective functions} during the observed time horizon has also been addressed in IRL. A multi-phase setting similar to the Brachistochrone example \eqref{eq_bracswitchB} above was considered in \cite{jin2019inverse}. Extending this to a general gIOC setting will involve using theory and algorithms from mixed-integer optimal control \cite{Sager2012-vd}.

\ereview

\sreview
\subsubsection{Identifiability and System Excitation} \label{sec_identifiability}

The extension of identifiability concepts to inverse optimal control presents additional layers of complexity beyond traditional system identification. In this context, we aim not only to identify system parameters but also to infer the objective function that explains observed trajectories as optimal solutions to a control problem. For nonlinear ordinary differential equations, this involves establishing conditions under which different objective functions lead to distinguishably different optimal trajectories, i.e. a form of structural identifiability specific to the inverse optimization framework. The challenge is compounded by the fact that optimal trajectories must satisfy necessary conditions from Pontryagin's Maximum Principle, leading to a two-point boundary value problem where both state and costate variables play crucial roles. Practical identifiability in this context must consider not only measurement noise and sampling limitations but also the possibility that observed trajectories may be only approximately optimal, either due to biological variability or because the true objective function includes terms or constraints not considered in our model class. Furthermore, the nonlinear nature of both the dynamics and the optimization problem can create multiple local optima \cite{Tsiantis2018,Tsiantis2020-nf}, making it difficult to distinguish between different objective functions that might produce similar trajectories. This necessitates careful consideration of both the model structure and the experimental design to ensure that the observed data contains sufficient information to discriminate between competing hypotheses about the underlying optimization principles governing the biological system.

\textbf{Persistent excitation} is a fundamental concept in system identification that characterizes the richness of input signals required to ensure identifiability \cite{Glad1990}. An input signal is persistently exciting when it contains sufficient frequency content to excite all relevant dynamic modes of the system, making them observable in the output data. For linear systems, this is typically expressed through conditions on the spectral content of the input signal or through properties of the information matrix, requiring it to be positive definite.

The role of persistent excitation is to provide sufficient conditions for parameter convergence in identification algorithms, ensuring that the experimental data contains enough information to distinguish between different possible model parameters and structures \cite{Angulo2017}. Without adequate excitation, multiple parameter sets might explain the observed input-output behavior equally well. While persistent excitation requirements are well-understood for linear systems and can be satisfied by signals such as pseudo-random binary sequences or multi-sine signals, the concept becomes more complex for nonlinear systems, where the required excitation may depend on the operating point and specific nonlinearities involved. This concept bridges the gap between structural and practical identifiability, and is particularly crucial in adaptive control and online identification schemes.

The relationship between persistent excitation, \textbf{statistical consistency}, and identifiability forms a fundamental framework in system identification theory \cite{ljung1999system}. Persistent excitation ensures that input signals provide sufficient information content, acting as a prerequisite for identifiability. Statistical consistency connects these concepts to the asymptotic behavior of parameter estimates. When an identification method is statistically consistent and the input is persistently exciting, the parameter estimates converge to their true values as the amount of data increases. This requires both structural identifiability (the theoretical possibility of unique parameter identification) and persistent excitation (sufficient information content in the data). The quality of persistent excitation affects the rate of convergence: stronger excitation generally leads to faster convergence and better numerical conditioning of the estimation problem. However, while persistent excitation and structural identifiability are necessary conditions for statistical consistency, they are not always sufficient, particularly in nonlinear systems where issues like local minima and parameter correlation can complicate the relationship between these properties \cite{Tsiantis2018}. In practice, these theoretical relationships guide experimental design and help determine how much data is needed to achieve desired estimation accuracy. While this theoretical framework is well-established for engineered systems where inputs can be precisely controlled, its application to biological systems presents unique challenges that we believe warrant careful consideration.

In molecular and cellular biology, achieving persistent excitation is particularly challenging, and in many cases, practically impossible. Biological systems are characterized by complex, nonlinear dynamics with multiple feedback loops, context-dependent responses to stimuli, and inherent stochasticity in cellular processes. These systems typically involve multiple interconnected regulatory networks and homeostatic mechanisms that actively resist perturbations. Furthermore, many system components have limited experimental accessibility, making comprehensive system identification particularly challenging \cite{Villaverde2014}.

While modern experimental techniques such as periodic drug administration, optogenetic stimulation, or carefully designed perturbation experiments can help probe system dynamics \cite{Caringella2023}, ensuring complete persistent excitation of all relevant modes remains an open challenge. Moreover, attempts to achieve persistent excitation might push the system outside its normal operating regime, potentially triggering compensatory mechanisms or stress responses that alter the very dynamics we aim to study. It should be noted that the role of noise and stochasticity is crucial, particularly in biological systems where variability arises both from measurement uncertainty and inherent cellular processes. 

Overall, identifiability and statistical consistency will play key roles in practical applications. Specifically, multiple experiments increasing data quantity and quality increases the information content and can help overcome these challenges and the conditions under which the inverse problem can be solved despite the practical limitations in achieving persistent excitation. Optimum experimental design (or active learning) is a systematic approach for obtaining useful data. Surveys of the design of optimal experiments in biological processes can be found in \cite{Abt2018-ct,Braniff2018-aq}. Early works in cell signaling are \cite{Balsa-Canto2008-lc,schenkendorf2009optimal}, and for the first time with experimental validation, \cite{Bandara2009-aj}.

Open questions in \textbf{optimal experimental design} (OED) for biological systems revolve around moving beyond traditional parameter refinement to instead handle the inherent uncertainty in these systems. This involves creating methods to distinguish between competing mechanistic models and extract maximum predictive value even from models known to be structurally non-identifiable \cite{Balsa-Canto2008-lc,Busetto2013,Treloar2022}. The question of whether an observable should be measured at a given time or not results in another layer of mixed-integer optimal control. Compare \cite{Sager2013-rn} for a discussion of theoretical properties and numerical methods. 

Furthermore, new approaches are needed to define informational richness (the equivalent of persistence of excitation) for biologically feasible inputs, which are often sparse and constrained rather than continuous, and to design experiments that can simultaneously inform dynamics across multiple time scales. The ultimate goal is to develop adaptive, closed-loop OED frameworks where experiments are intelligently updated in real-time to most efficiently resolve these ambiguities, thus accelerating scientific discovery. Designing optimal experiments not only for parameter estimation but also for the concurrent discrimination of objectives, constraints, and models is completely open. 
\ereview

\begin{figure*}[b!!!]
\centering
  \includegraphics[width=\linewidth]{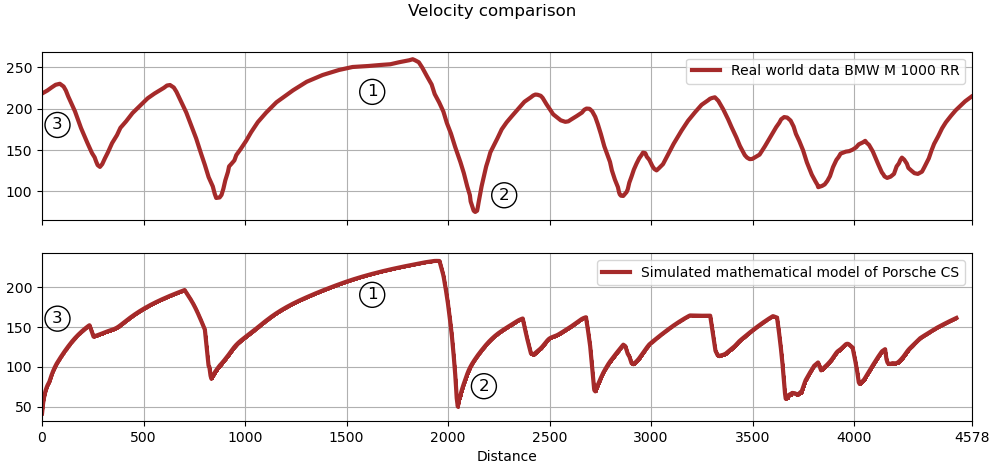}
\caption{Comparison of different near-time-optimal velocity profiles for human and mathematically optimized driving. It has been shown that human professionals drive almost optimally in races \cite{dal2018minimum}. Here, complementing the optimal controls shown in Figure~\ref{fig_carAdvanced}, velocity profiles for the Hockenheim ring are displayed. A comparison of the curves for a professional motorbike driver provided by the magazine \textit{PS}, issue 9/2022, page 45, on the top and the optimal solution from \cite{Kehrle2011} for a mathematical model on the bottom shows many similarities. Of particular interest is \myball{1} a comparison of the increased velocities resulting from full acceleration resulting in very similar shapes, although the maximum acceleration and absolute velocity are different. One observes also the impact of \myball{2} active path constraints forcing the drivers to brake heavily to stay on the racing track and of \myball{3} initial conditions (the simulated car starts with $v=0$ in round 1 unlike the motorbike). Intuitively and visually, even laymen can identify the underlying optimality principle, which comprises the minimal time objective function. This principle is also significantly influenced by the sharp curves leading to sharp declines in velocity and in the first part of the track by the initial conditions. Note that the corresponding optimal gear choices shown in Figure~\ref{fig_carAdvanced} are linked via engine speed constraints to the velocity of the car. Thus it appears plausible that such an inference would also be possible based on observed gear choices instead of observed velocities. We view this as an encouraging example for gIOC tasks in cell biology. However, the question of which observations and how many are needed to uniquely identify objective function and constraints in a gIOC problem remains open. It seems highly nontrivial to develop algorithms to systematically identify optimality principles in the general case.
}
\label{fig_carComparison}
\end{figure*}

\subsubsection{Examples of gIOC} \label{sec_gIOCexamples}

Although first results like in \cite{Tsiantis2018} have been obtained for IOC problems in biology, there is currently no numerical solver for gIOC problems available. Therefore we provided examples in Figures~\ref{fig_bracExamples} and \ref{fig_metabolicExamples} that give a visual illustration of the gIOC task and associated challenges.

Another illustrative example is given in Figure~\ref{fig_carComparison}. It shows a comparison between observed human and simulated mathematical race driving. The similarities in the velocity profiles are clearly identifiable, even for laymen. The velocity profile of the professional motorbike driver is periodic while the optimization assumed round 1 with initial zero velocity as an initial condition. The motorbike has different maxima of acceleration and velocity compared to the model parameters $p$ of the Porsche CS model used by \cite{Kehrle2011}, and take different times to cover the complete Hockenheim track (not shown). Nevertheless, the velocity profile plotted over covered distance on the Hockenheim track shown in Figure~\ref{fig_carAdvanced} is similar in shape to the optimal forward solution of the mathematical model. This allows for visually identifying the intrinsic optimality principle consisting of minimum time driving, initial conditions, and active path constraints. 

We see this as an \textbf{encouragement} that, despite the open theoretical questions on the quality and quantity of data $\eta$ required to uniquely identify optimal control problems and the difficulties to implement numerical algorithms to do this systematically in the general case, and despite the obvious mismatch between mathematical model and real world process, also in other biological applications a practical inference of optimality principles should be possible.

\subsubsection{Summary}

The gIOC problem class has high practical relevance with outstanding chances of enabling novel and impactful insights in mathematical biology. At the same time, it has not yet been receiving much attention from the mathematical community at large. We expect that the interesting structure, featuring a data fit problem on the outer and a nonconvex control problem with many unknowns on the inner level, will stimulate cross-discipline research. Example topics are structural properties of the resulting constrained optimization problems, identifiability and observability issues, strategies for designing experiments that yield high-quality data, the development, implementation, and analysis of efficient algorithms, as well as links to symbolic regression and concepts in RL and IRL. We are not aware of any method or solver able to solve it in full generality. One reason for this is that many relevant aspects have only been investigated separately and in different mathematical communities, focussing either on symbolic regression, nonlinear / bi-level / mixed-integer / robust / multi-objective optimization, numerics, optimal control, systems theory, observability and identifiability, mathematical biology, experimental design, or the various research areas associated with machine learning. 
\sreview
Currently, the most active community in method development seems to be related to robotics and reinforcement learning as discussed in Sections~\ref{sec_irl} and \ref{sec_bilevel}.
\ereview
%
We are optimistic that with the right interdisciplinary approach, including development of the necessary models, algorithms, theory, and solvers, gIOC can become an enabling technology \sreview also in molecular and cell biology.\ereview

\section{Conclusion} \label{sec_discussion}

We aimed to present a compelling argument that optimality principles are ubiquitous in biology. This may be due to a mixture of underlying physical and chemical optimality principles and of evolutionary and training effects. We surveyed and discussed controversial points and possible counter-arguments, asserting that it is worthwhile to find such principles independent of their origin and evolutionary course: to enable knowledge gain in biology and new control concepts in bio-engineering.

The optimality of structures, shapes, and dynamic behavior in biology is supported by a wealth of studies across various scales, ranging from molecular to ecosystem levels. However, the conventional practice of invoking an optimality principle without a clear definition and quantification of the optimization target has faced criticism. Retroactively introducing additional constraints to justify a specific outcome may compromise the robustness of the model.
Therefore we propose an inverse approach to infer optimality principles from data. We present a generalized inverse optimal control framework, which is based on the following elements:

\begin{itemize}
\setlength{\itemsep}{-2pt}
 \item balancing multiple objectives: biological systems often strive for a trade-off between goals.
 \item nested hierarchy: objectives can be nested, operating at different scales and switching over time, all contributing to overall inclusive fitness maximization.
 \item dynamic constraints: real-world organisms operate within limitations that may change (becoming active / inactive) over time.
 \item robustness as a goal: organisms often prioritize resilience to environmental fluctuations.
 \item individual variation: objectives may differ slightly between individual organisms within a population.
 \item modeling uncertainty: our understanding and prior knowledge of biological systems is incomplete.
\end{itemize} 

We discussed and illustrated challenges like the ill-posedness of the problem, considering both structural and practical perspectives. Moreover, the development of efficient and robust numerical methods will be essential for gIOC, encompassing optimum experimental design, bi-level optimization, optimal control, symbolic regression, dynamic game theory, and reinforcement learning.

Although our main focus here is on molecular systems biology, it's important to emphasize that the applicability of our framework is not confined to this level alone, and it can be expanded to encompass other tiers of biological organization. For example, co-evolutionary phenomena could be integrated in \eqref{eq_generalIOC} using concepts of game theory \cite{Smith1982,sigmund1999evolutionary,Schuster2008,hummert2014evolutionary,garde2020modelling}. Therefore, looking ahead, we are confident that the complexities of these scenarios can be effectively captured by further generalizing our approach using inverse dynamic games \cite{molloy2022inverse}. 

We believe that achieving this goal requires an interdisciplinary effort. Thereby, we can ensure that the principles we uncover are not merely speculative but firmly rooted in experimental evidence. This approach may foster theoretical understanding and facilitate the application of the optimality principle in forward optimal control applications across biomedicine, biotechnology, and agriculture.

\printcredits


\noindent
\textbf{Acknowledgements.}
%
JRB acknowledges  financialsupport from grant PID2020-117271RB-C22 (BIODYNAMICS) funded by MCIN/AEI/10.13039/501100011033, 
from grant PID2023-146275NB-C22 (DYNAMO-bio) funded by MICIU/AEI/ 10.13039/501100011033 and ERDF/EU, and from grant CSIC PIE 202470E108 (LARGO).
SS acknowledges financial support from the research initiative ``SmartProSys: Intelligent Process Systems for the Sustainable Production of Chemicals'' funded by the Ministry for Science, Energy, Climate Protection and the Environment of the State of Saxony-Anhalt and from the German Research Foundation DFG via grant 314838170, GRK 2297 MathCoRe.\\
We thank 
Eva Balsa-Canto,  
Steffen Klamt,
Mariko Okada,
Andreas Schuppert, and
Alejandro F. Villaverde 
for stimulating discussions and feedback.

\bibliographystyle{elsarticle-num}

\bibliography{bib/ioc-bibliography}



\end{document}